\documentclass[12pt]{article}

\usepackage[letterpaper, portrait, margin=1.25in]{geometry}
\usepackage[utf8]{inputenc}

\usepackage{changes}
\usepackage{bm}
\usepackage{aliascnt}
\usepackage{amsmath,amssymb,amsthm}
\usepackage{setspace}
\usepackage{booktabs}
\usepackage{xcolor}
\usepackage{soul}
\usepackage{tikz}
\usetikzlibrary{arrows.meta}
\usetikzlibrary{positioning,arrows.meta,decorations.pathreplacing}

\usepackage{natbib}
\usepackage{xurl}
\usepackage{float}

\usepackage{hyperref}
\hypersetup{
    colorlinks=true,
    linkcolor=blue,
    urlcolor=blue,
    citecolor=blue
}

\usepackage{algorithm}
\usepackage{algpseudocode}

\theoremstyle{plain}
\newtheorem{theorem}{Theorem}
\newtheorem{proposition}{Proposition}
\newtheorem{lemma}{Lemma}
\newtheorem{definition}{Definition}

\newtheorem{example}{Example}

\newtheoremstyle{remarkbold}
  {}
  {}
  {\normalfont}
  {}
  {\bfseries}
  {.}
  {.5em}
  {}

\theoremstyle{remarkbold}
\newtheorem{remark}{Remark}

\definecolor{frontierblue}{RGB}{8,22,103}
\definecolor{thresholdred}{RGB}{112,8,18}

\title{Reserve Systems with Match-Specific Beneficiaries\thanks{An earlier version of this paper has been circulated under the title ``Reserve System with Beneficiary-Share Guarantee.'' We wish to thank the Editor (Faruk Gul), two anonymous reviewers, In\'{a}cio B\'{o}, Olivier Bochet, Yeon-Koo Che, Julien Combe, David Delacr\'{e}taz, Di Feng, Guillaume Haeringer, Kenzo Imamura, Aditya Kuvalekar, SangMok Lee, Mengling Li, Vikram Manjunath, Franz Ostrizek, Eduardo Perez-Richet, Marek Pycia, Olivier Tercieux, M. Utku \"{U}nver, Sonal Yadav, and M. Bumin Yenmez for helpful comments, as well as participants at GBA Market Design Workshop 2025 at University of Macau,  Sciences Po D1 Presentation, PSE ARGET, IES Workshop on Mechanism Design at SWUFE, ISI Delhi, WZB Berlin Workshop on Matching and Market Design, Ashoka University, and Shiv Nadar University.}}

\author{Yuan Gao\thanks{Washington University in St. Louis, St. Louis 63130, United States. Email: ygao1@wustl.edu} 
\and Xi Jin\thanks{Sciences Po, Paris 75007, France. Email: xi.jin1@sciencespo.fr}
\and Manshu Khanna\thanks{Peking University HSBC Business School, Shenzhen 518055, China. Email: manshu@phbs.pku.edu.cn}}

\begin{document}
 
\maketitle

\begin{abstract}
We study two-sided matching problems in which the designer regards certain participant--institution matches as socially desirable (‘beneficiary matches’) and seeks to promote such matches. This objective can conflict with maximizing the total number of matches. We introduce minimal cycles to characterize the complete non-domination frontier, where each point represents an allocation that cannot increase beneficary matches without sacrificing total matches.
Our main results are (i) the frontier is concave, so each additional match costs weakly more beneficiary matches than the last, (ii) traversing from maximum total matches to maximum beneficiary matches on the frontier reduces total matches by at most half of the maximum total, (iii) the Repeated Hungarian Algorithm computes the entire frontier in polynomial time, and (iv) mechanisms that approximately satisfy a percentage requirement of beneficiary matches on the frontier can respect priority orderings and elicit eligibility in a strategy-proof manner, but no such mechanism is path-independent. These results enable rigorous evaluation of policies that promote beneficiary matches across diverse allocation contexts.

\end{abstract}

\bigskip
\noindent\textbf{JEL Classification:} C78, D47, D61, D63, I28 

\bigskip
\noindent\textbf{Keywords:} Matching with constraints, market design, reserve system, daycare assignment, affirmative action

\newpage
\doublespacing

\section{Introduction}
In many matching markets, policymakers pursue objectives whose social benefits are not fully reflected in participants’ preferences. Incorporating such objectives therefore requires additional structures in the matching process.

A common approach is to reserve positions for designated groups of participants, such as minority students, disadvantaged applicants, or essential workers. The underlying objectives are \emph{participant-specific}: the goal is to promote the representation of certain groups, as in affirmative action and reserve systems \citep{hafalir2013,kojima2012,sonmez2022yenmez,pathak2024,aygun2017,aygun2020,aygun2022ms,turhan2024}. Throughout, we say that a match is \emph{beneficiary} if
the designer deems it socially desirable. Under a participant-specific objective, every match involving a member of the designated group is beneficiary, regardless of the institution to which the participant is
assigned.

More generally, the objective can be \emph{match-specific}: whether a match is beneficiary depends on the participant--institution pair, not on the identity of the participant alone. Many policies have this structure. Municipalities throughout Europe give priority to children seeking admission to nearby day care centers or schools,\footnote{See, for example, Estonia's Preschool Child Care Institutions Act (\url{https://www.riigiteataja.ee/en/akt/518122017001}), Sweden's proximity principle (\url{https://utbildningsguiden.skolverket.se/}), and Catalonia's school admission policy (\url{https://preinscripcio.gencat.cat/ca/estudis/infantil-3-6/informat/criteris-assignacio/index.html}).} civil servants may be preferentially assigned to states away from their home state,\footnote{See the Government of India's IAS Cadre Allocation Policy (\url{https://drive.google.com/file/d/1GUQt3MnnY1D6gMc2fL1BvTUj1HseB-lf/view}) and \citet{bo2026visiblyfairmechanisms}.} universities allocate parking according to employees' primary work locations,\footnote{See UCSF Parking Policy 200-17 (\url{https://policies.ucsf.edu/policy/200-17}).} and integration-driven school districts grant priority to disadvantaged students specifically at high-performing schools.{\interfootnotelinepenalty=10000
\footnote{See the SFUSD tiebreaker policy
(\url{https://www.sfusd.edu/schools/enroll/student-assignment-policy/tiebreakers})
and the LAUSD Magnet Priority Point System
(\url{https://choices.lausd.net/magnet}).}} In each of these settings, the same participant may form a beneficiary match at one institution but not at another. A similar structure arises in hospital operations, where assigning a patient to a generalist is acceptable but assigning the patient to the designated specialty service is socially preferable \citep{song2020capacity,dong2018offservice}.\footnote{See NHS e-Referral Service guidance on named clinician referrals (\url{https://digital.nhs.uk/services/e-referral-service/document-library/named-clinician-guidance}).} Such match-specific objectives have recently begun to receive attention in the theoretical matching literature; see, for example, \citet{evren2023reserve}.

Whether participant-specific or match-specific, the objective can be incorporated by constraining the number of beneficiary matches. Such a constraint restricts the set of admissible matchings and may therefore conflict with other objectives. Our focus is its effect on matching capacity: the extent to which promoting beneficiary matches reduces the total number of matches.

For participant-specific objectives, promoting beneficiary matches entails no loss of capacity. For match-specific objectives, by contrast, a loss generally arises (\autoref{rmk:comparison with type approach}) and can be as large as one half of the maximum number of matches (\autoref{conflict}). Thus, even setting aside fairness and efficiency, the structure of the objective alone can reduce the number of participants who are matched. This concern is acute in applications such as day care allocation \citep{sun2022daycarematchingjapantransfers} and vaccine allocation \citep{evren2023reserve}, where the relevant objectives are typically match-specific. This paper characterizes the trade-off between matching capacity and match-specific objectives, providing a framework with which policymakers can weigh the two goals.

We develop the framework through a leading application: assigning children to day care centers.\footnote{See footnote 1 for examples from Estonia, Sweden and Catalonia.} 
Each center has limited capacity, and its \emph{eligibility set} consists of the children who may be assigned to it. Eligibility can reflect both institutional admissibility and family-side restrictions on feasible assignments; in the main model, we take the resulting pair-specific eligibility relation as given.
Among the children eligible at each center, the designer designates a subset as \emph{beneficiaries}---for example,  those living within the center's catchment area---whose placement there the social objective seeks to promote. Beneficiary status is match-specific: a child living near center \(c\) but far from \(c'\) is a beneficiary for \(c\) and not for \(c'\). That is to say, it is a property of the (child, center) pair rather than of the child alone. The designer values proximity for reasons external to family preferences---congestion, district-level spending, neighborhood cohesion---but promoting it is costly in terms of capacity utilization.

Insisting that every child be placed at a nearby center is too rigid: it risks leaving children unmatched altogether, the worst outcome. ``Didn't get a slot in day care. Drop dead, Japan!!!'', a viral
2016 blog post, captures the salience of children left entirely
unplaced \citep{Osaki2016}.\footnote{The concern is particularly
important when facing shortages. For instance, there is a shortfall of
430,000 day care spots in Germany as of November 2024 (\href{https://www.dw.com/en/germany-banks-on-immigrants-to-deal-with-its-child-care-crisis-v3/a-70795228}{DW, 2024}).
\citet{afacan2023assignment} examine this objective in a school
choice setting.} Ignoring proximity, however, abandons the social objective. Existing approaches typically resolve this tension by lexicographically prioritizing one objective over the other \citep{evren2023reserve}. Between these two extremes lies a continuum of compromises, which we formalize as the \emph{non-domination frontier}, the set of outcomes for which no alternative yields both more total matches (\(e\)) and more beneficiary matches (\(b\)) at once (\autoref{def of frontier}). The frontier identifies, at each level of total matches, the exact rate at which beneficiary matches must be surrendered to increase match quantity.

We characterize the $(e,b)$ pairs on the frontier through a new structure we term \emph{minimal cycles} (\autoref{def:minimal cycle}). The graph-and-cycle apparatus itself is widely used in matching theory. There have been top trading cycles \citep{ShapleyScarf1974,AbdulkadirogluSonmez2003}, stable improvement cycles \citep{ErdilErgin2008}, and other related cycle-based constructions \citep{KamadaKojima2025iBF}. Unlike these cycles, whose properties arise from greedy edge choices based on participants' priorities or preferences, our minimal cycles can replace a beneficiary match with a merely eligible match, exhibiting a non-greedy feature at the edge level. Instead, they are greedy at the cycle-level: the choice of cycle requires the least beneficiary loss, which aggregates the per-edge losses along the cycle (\autoref{def:minimal cycle}). Iterating minimal cycles traces the entire non-domination frontier. The frontier is concave, exhibiting diminishing returns, so each additional eligible match costs weakly more beneficiary matches than the last (\autoref{slope of the frontier}). Moreover, prioritizing beneficiaries never reduces matching capacity by more than half: moving along the frontier from the eligibility-maximal point to the beneficiary-maximal point costs at most 50\% of total matches, and this bound is tight (\autoref{Upper bond}). These properties hold for every instance, so a policymaker can assess the fundamental trade-offs before collecting any market-specific data. We then develop the Repeated Hungarian Algorithm to compute every frontier point in $O(n^{4})$ time (\autoref{alg:RHA}, \autoref{Property of RHA}), rendering the characterization operational.

In practice, policymakers who require explicit guidance often mandate a target ex ante: a beneficiary share, \(\beta^{*}\), requiring beneficiary matches to constitute at least a specified fraction of all matches \citep{dur2018, abdulkadiroglu2005nyc}. Examples include Boston's 50\% walk-zone policy or Karnataka's 75\% local-taluk reservation.\footnote{See \url{https://www.deccanherald.com/amp/story/india/karnataka/75-local-quota-rule-in-kreis-schools-disappoints-students-from-taluks-with-fewer-institutions-3551461}} The frontier presents the entire menu of such possibilities (\autoref{figure intro}).  We show, however, that enforcing exact compliance with a beneficiary-share target can produce dominated outcomes: a matching that meets the target with equality may lie strictly below the frontier, in which case total matches could be increased without reducing beneficiary matches (\autoref{beta threshold}). This motivates our approach of ``approximate'' compliance, which selects the frontier matching whose beneficiary share weakly exceeds the target by the smallest margin when possible and a max-bene-then-max-elig matching otherwise. When such approximation is required, the resulting frontier matching dominates every matching that meets the target exactly, improving one objective without worsening the other (\autoref{justification of approximation}, \autoref{figure intro}).

\begin{figure}[t]
    \begin{tikzpicture}[
  x=0.1cm,
  y=0.1cm,
  line cap=round,
  line join=round,
  every node/.style={font=\small},
  ray/.style={dashed, dash pattern=on 4pt off 3pt, line width=1.05pt},
  arrowtip/.style={-{Stealth[length=3.4mm,width=2.6mm]}},
]

  \draw[-{Stealth[length=4mm,width=3mm]}, line width=0.9pt] (0,0) -- (110,0);
  \draw[-{Stealth[length=4mm,width=3mm]}, line width=0.9pt] (0,0) -- (0,80);

  \foreach \x in {20,40,60,80,100} {
    \draw[line width=0.7pt] (\x,-1) -- (\x,1);
    \node[below=5pt] at (\x,0) {\x};
  }
  \foreach \y in {10,20,30,40,50,60,70} {
    \draw[line width=0.7pt] (-1,\y) -- (1,\y);
    \node[left=5pt] at (0,\y) {\y};
  }
  \node[below left=4pt and 2pt] at (0,0) {0};

  \node[below=22pt] at (55,0) {\normalsize total matches $e$};
  \node[rotate=90] at (-11,40) {\normalsize beneficiary matches $b$};

  \draw[ray, arrowtip, black] (0,0) -- (100,20)
    node[pos=0.62, sloped, above=4pt, text=black] {$\beta \ge 20\%$};

  \draw[ray, arrowtip, thresholdred] (0,0) -- (94,47)
    node[pos=0.55, sloped, above=4pt, text=thresholdred]
      {$\beta \ge 50\%$ (Boston)};

  \draw[ray, frontierblue] (0,0) -- (84,63)
    node[pos=0.5, sloped, above=4pt, text=frontierblue]
      {$\beta \ge 75\%$ (Karnataka)};
  \draw[arrowtip, thresholdred, line width=1.15pt] (84,63) -- (87,68);

  \draw[frontierblue, line width=1.2pt, smooth, tension=0.7]
    plot coordinates {
      (86,71) (87,68) (90,59) (92,53) (94,47) (96,40) (98,30) (100,20)
    };

  \foreach \p in {(86,71),(90,59),(92,53),(96,40),(98,30),(100,20)} {
    \fill[frontierblue] \p circle[radius=2.1pt];
  }

  \fill[thresholdred] (84,63) circle[radius=2.3pt]; 
  \fill[thresholdred] (87,68) circle[radius=2.4pt]; 
  \fill[thresholdred] (94,47) circle[radius=2.4pt]; 

  \node[text=frontierblue, above=2pt]                at (86,71) {$\mu_{BE}\;(86,71)$};
  \node[text=thresholdred, above left=2pt and 1pt, yshift=-9pt]    at (84,63) {$(84,\,63)$};
  \node[text=thresholdred, right=3pt]                at (87,68) {$(87,\,68)$};
  \node[text=thresholdred, right=3pt]                at (94,47) {$(94,\,47)$};
  \node[text=frontierblue, above right=1pt and 1pt, yshift=-9pt]  at (100,20) {$(100,\,20)$};
  \node[text=frontierblue, below right=1pt and 1pt]  at (100,20)
    {Evren (2023), $\mu_{EB}$};

\end{tikzpicture}
\caption{\textbf{Beneficiary-share policy along the non-domination frontier.}
The blue curve is the non-domination frontier.  A target \(\beta^{*}\) corresponds to the ray \(b=\beta^{*}e\); points on or above the ray satisfy the target. The eligibility-maximal frontier point corresponding to \(\mu_{EB}\), namely \((100,20)\), attains a \(20\%\) beneficiary share, and Boston's \(50\%\) target is met exactly at the frontier point \((94,47)\). Exact compliance with Karnataka's \(75\%\) target at \((84,63)\) lies below the frontier and is dominated by \((87,68)\), which attains more total matches and more beneficiary matches while approximately satisfying the
target.}
\label{figure intro}
\end{figure}

The framework extends to center-specific priority orderings over parents, while preserving approximate beneficiary-share compliance on the frontier (\autoref{Respecting priority}).\footnote{A matching respects these priorities if no unmatched parent has higher priority at a center than a parent assigned there.} 
We also show that parent-side eligibility information can be elicited in a strategy-proof manner while preserving approximate beneficiary-share compliance on the frontier (\autoref{Strategy-proof}).
By contrast, no mechanism that respects the beneficiary-share guarantee approximately on the frontier can satisfy path-independence---the property that the final choice remains the same whether we consider all candidates at once or first select from subgroups and then choose from those selected candidates. This impossibility holds even in the absence of priorities, revealing a fundamental tension between non-domination and procedural consistency
(\autoref{thrm PI imposible}).

The analysis to this point requires no information about how families rank the centers at which they are eligible.  Our emphasis on avoiding wasteful assignments by placing as many children as possible is consistent with related matching environments in which preferences among acceptable alternatives are coarse or secondary \citep{aziz2025strategyproof,JARAMILLO20121913,erdil2017two}.
Next, we introduce strict preferences over centers to study how frontier matchings interact with individual welfare and fairness.

Once strict preferences over centers are introduced, non-domination from the designer's perspective need not imply Pareto efficiency or fairness for parents. This tension echoes the broader difficulty of reconciling non-wastefulness, allocation constraints, and fairness \citep{KamadaKojima2024,fragiadakis2017}.\footnote{\citet{KamadaKojima2024}, studying the non-existence of stable matchings, explicitly contrast relaxing non-wastefulness while preserving fairness with the alternative of maintaining non-wastefulness while relaxing fairness.} We therefore ask which welfare properties can be recovered conditional on remaining on the frontier. We show that a constrained efficient matching at any frontier point $(e,b)$ is computable in polynomial time (\autoref{prop:ce lp}). Moreover, when every parent ranks the centers at which she is a beneficiary above those at which she is not, every frontier point admits a Pareto efficient matching (\autoref{aligned efficiency}). Fairness, however, may fail even under this preference restriction (\autoref{nd not fair}).

\paragraph{Related Literature.} Our paper studies a match-specific beneficiary system, which can be seen as an extension of the standard participant-specific reserve system. The theory of reserve systems emerged from \cite{hafalir2013}, who showed how reserve categories implement affirmative action in school choice, addressing limitations of quotas identified by \cite{kojima2012}. \cite{sonmez2022yenmez} extended this to overlapping reserves and the meritorious horizontal choice rule, and \cite{pathak2024} showed how reserves balance competing ethical claims in pandemic medical-resource allocation. Most recently, \cite{evren2023reserve} studied the match-specific beneficiary system in the vaccine allocation problem, and showed how to find a lexicographic solution in such a system (\autoref{remarkmaxinmax}), which is an endpoint of the non-domination frontier in this paper. We extend this line of research by characterizing the full non-domination frontier in the match-specific beneficiary system. 

Real-world implementations, from Boston's walk zones \citep{dur2018} to H-1B visa reform \citep{pathak2020}, underscore the value of understanding feasible outcomes before deployment. \citet{pathak2023reversing} show experimentally that 40\% of participants conflate minimum guarantees with over-and-above allocations. Experience with Indian colleges' affirmative action policies has prompted further advances: \cite{turhan2024} extended gradual matching to complex affirmative action settings, while \cite{aygun2017}, \cite{aygun2020}, and \cite{aygun2022ms} examined large-scale implementations, dynamic reserves, and de-reservation through choice rules. Our frontier directly quantifies the trade-offs these settings face. The work also connects to choice-rule design: characterizing substitutability \citep{aygun2013}, affirmative-action schemes \citep{echenique2015, imamura2025meritocracy}, slot-specific priorities \citep{kominers2016}, and diverse policy objectives \citep{dougan2025market, yokote2023representation}. Where \citet{dougan2025market} emphasize path-independence \citep{chambers2017choice, plott1973}, we show that pursuing non-domination under approximate share guarantees necessarily violates substitutability, and hence path-independence.

The paper is organized as follows. Section~2 presents the model and the frontier concept. Section~3 characterizes the frontier and introduces the Repeated Hungarian Algorithm with its complexity analysis. Section~4 examines how non-domination interacts with other key allocation criteria. Section~5 concludes. Omitted proofs appear in the appendix.

\section{Model}

We consider a two-sided matching problem with a finite set of centers $\mathcal{C} = \{c_1, c_2, \ldots\}$ on one side of the market and a finite set of parents $\mathcal{P} = \{p_1, p_2, \ldots\}$ on the other. The designer's preferences over matches are two-tiered, captured by an eligibility structure and a beneficiary structure. For each center $c \in \mathcal{C}$, the set $E_c \subseteq \mathcal{P}$ denotes the parents who are mutually acceptable to $c$; we write $\mathcal{E} = (E_c)_{c \in \mathcal{C}}$. Within each eligible set, a subset $B_c \subseteq E_c$ identifies the parents whose assignment to $c$ the designer deems socially desirable; we write $\mathcal{B} = (B_c)_{c \in \mathcal{C}}$. Each center $c$ has a quota $q_c \in \mathbb{N}$ specifying the maximum number of parents it can admit, and we write $\mathbf{q} = (q_c)_{c \in \mathcal{C}}$.

We call the tuple
$(\mathcal{C},\mathcal{P},\textbf{q},\mathcal{E},\mathcal{B})$
an \textbf{instance}. For any instance, a correspondence
$\mu:\mathcal{C}\cup\mathcal{P}\rightrightarrows
\mathcal{C}\cup\mathcal{P}$
is a \textbf{matching} if:
(i) each parent is matched to at most one center
($\mu(p)\subseteq\mathcal{C}$ with $|\mu(p)|\leq 1$ for all
$p\in\mathcal{P}$, where $\mu(p)=\emptyset$ denotes being unmatched);
(ii) each center is matched only to parents and respects its quota
($\mu(c)\subseteq\mathcal{P}$ with $|\mu(c)|\leq q_c$ for all
$c\in\mathcal{C}$); and
(iii) assignments are consistent
($\mu(p)=\{c\}$ if and only if $p\in\mu(c)$).
When unambiguous, we write $\mu(p)=c$ instead of $\mu(p)=\{c\}$.
A matching $\mu$ is \textbf{eligible} if
$\mu(p)=c\in\mathcal{C}\implies p\in E_c$.
We consider only eligible matchings unless stated otherwise.

For matching $\mu$, we define \textbf{beneficiary matches} as $B(\mu) = \{ (p, c) \mid p \in \mathcal{P}, c \in \mathcal{C}$, $p \in \mu(c) \cap B_c \}$, 
and for center $c$, $B_\mu(c) = \left\{ (p, c) \mid p \in \mu(c) \cap B_c \right\}$. 
Similarly, \textbf{eligible matches} are $E(\mu) = \left\{ (p, c) \mid p \in \mathcal{P}, c \in \mathcal{C}, p \in \mu(c)\cap E_c \right\}$, 
and $E_\mu(c) = \left\{ (p, c) \mid p \in \mu(c) \cap E_c \right\}$. When context is clear, we use these terms to refer to either the sets or their cardinalities, and use “total matches” interchangeably with “eligible matches.”
 
A \textbf{mechanism} \(\varphi\) is a function which takes an instance as input and outputs an eligible matching \(\mu\).\footnote{In Section~4, where a beneficiary-share guarantee is added to the primitives, a mechanism takes the augmented problem as input.}

\subsection{The Frontier}

Two competing objectives arise in allocation: maximizing eligible matches for broad access versus maximizing beneficiary matches for targeted priorities. To address this tension, \citet{evren2023reserve} proposed \textbf{max-in-max}, which lexicographically prioritizes objectives: selecting the matching with maximum beneficiary matches among those maximizing eligible matches. However, lexicographic maximization can entirely neglect one objective:

\begin{example}\label{conflict}
Consider parents $\mathcal{P}=\{p_1,p_2\}$ and centers $\mathcal{C}=\{c_1,c_2\}$ each with unit quota. Parent $p_1$ is eligible for both centers and a beneficiary of $c_2$; parent $p_2$ is eligible only for $c_2$. The unique eligible-maximizing matching assigns $\mu(p_1)=c_1$ and $\mu(p_2)=c_2$, yielding zero beneficiary matches.\footnote{Although parent $p_1$ is a beneficiary for $c_2$, her match to $c_1$ in the eligible-maximizing allocation does not count as a beneficiary match---beneficiary status only applies when matched to the designated center. In the day care setting, this captures the policy goal of placing children at centers within their own catchment area.
}\textsuperscript{,}\footnote{The cost of prioritizing total matches can be severe: our construction in \autoref{shorcoming of max in max} shows that maximizing eligible matches first can sacrifice arbitrarily many beneficiary matches---the loss grows unboundedly with market size.}
\end{example}

While lexicographic solutions prove problematic, the underlying principle remains appealing: conditional on eligible matches, maximize beneficiary matches; conditional on beneficiary matches, maximize eligible matches. We formalize this idea in the following definition.

\begin{definition}[Domination and Frontier]\label{def of frontier}
For any instance, a matching $\mu$ is \textbf{dominated} by a matching $\mu'$ if
\[
|E(\mu')| \geq |E(\mu)| 
\quad \text{and} \quad 
|B(\mu')| \geq |B(\mu)|,
\]
with at least one strict inequality.\footnote{Unlike the roommate problem domination in \citet{molis2024experimental}, where matchings dominate via blocking pair chains, our definition operates at the aggregate level.}
By definition, domination is a transitive relation. A matching is \textbf{undominated} if no other matching dominates it. A mechanism $\varphi$ is \textbf{undominated} if it always produces undominated matchings. The \textbf{non-domination frontier} is
\[
F
=
\left\{
\bigl(|E(\mu)|,|B(\mu)|\bigr)\in\mathbb{N}^{2}
:
\mu \text{ is undominated}
\right\}.
\]

\noindent An undominated matching $\mu$ \textbf{corresponds to} a point $(e,b)$ on the frontier $F$ if
\[
|E(\mu)|=e
\quad\text{and}\quad
|B(\mu)|=b.
\]
Importantly, the frontier is a set of attainable \emph{match-count pairs} $(e,b)$, rather than a set of matchings themselves. Thus, multiple distinct undominated matchings may correspond to the same frontier point.
\end{definition}

\begin{remark}\label{remarkmaxinmax}
    The max-in-max solution \citep{evren2023reserve} lies on the frontier but captures only one extreme point. 
\end{remark}

\begin{remark}\label{rmk:comparison with type approach}
    We emphasize that it is the match-specific nature of beneficiary status that gives rise to a rich non-domination frontier. In \autoref{conflict}, beneficiary status is determined by the parent--center pair rather than by the parent's identity alone: parent \(p_1\) is a beneficiary for center \(c_2\) but not for \(c_1\). For a formal counterpart, see \autoref{common bene set} in the Appendix which shows that when beneficiary status depends only on the parent---an assumption common in the reserve-system literature---the frontier degenerates to a single point.
\end{remark}

Non-domination provides both a normative criterion and a practical framework for understanding trade-offs between eligible and beneficiary matches.  The non-domination frontier, combined with a ray at the angle corresponding to a target share (introduced in Section~\ref{sec:beneguar}), delimits policy-required allocations. Section 3 characterizes the frontier and develops algorithms for its computation. Section 4 examines how non-domination interacts with other key allocation criteria.

\section{Characterization and Computation of the Frontier}\label{sec: characterization of frontier}
We characterize the non-domination frontier and present a polynomial-time algorithm for its computation.

\subsection{Cycle Characterization}
We provide a constructive characterization of the \((e,b)\) pairs on the frontier through a new structure we term \emph{minimal cycles}  (\autoref{def:minimal cycle}).

Without loss of generality, assume unit quotas, allowing us to write $\mu(c)=\{p\}$ as $\mu(c)=p$; likewise, for parents we write $\mu(p)=c$ rather than $\mu(p)=\{c\}$ (with slight abuse of notation).
We call unit-quota centers seats. 
For each undominated matching $\mu$, define a new graph, termed \textbf{associated directed graph}, $G_{\mu}(\mathcal{P}, \mathcal{C}, Eg_{\mu})$, which shows the eligibility relation and the current matching $\mu$. It is a directed bipartite graph with parents $\mathcal{P}$ and centers $\mathcal{C}$ as vertex sets. Its edge set $Eg_{\mu}$ contains: an edge from $p$ to $c$ if $p \in E_c$ and $\mu(p) \neq c$; an edge from $c$ to $p$ if $\mu(c) = p$ or $\mu(p) = \mu(c) = \emptyset$.

A \textbf{cycle} in $G_\mu$ is a sequence $\textbf{c} = (p_1, c_1,\ldots,p_n,c_n,p_1)$ where consecutive vertices are connected by directed edges. A cycle represents an eligible reallocation. A cycle $\textbf{c}$ is \textbf{unit eligibility improving} if $p_1$ and $c_n$ are the unique unmatched parent and seat under $\mu$ in $\textbf{c}$, respectively. 

\textbf{Applying} a unit eligibility improving cycle $\mathbf{c}$ (in $G_{\mu}$) to $\mu$ reassigns each parent $p_k$ to center $c_k$, yielding matching $\mathbf{c}(\mu)$ where:

$$ \textbf{c}(\mu)(p)=\begin{cases}
    \mu(p) & \text{if } p\notin \mathbf{c} \\ c_{k} & \text{if } p=p_k\in \textbf{c} \text{ for some }k.  \end{cases}$$ The resulting matching remains eligible by construction. Consider the following example:

\begin{example}
Consider parents $ \mathcal{P} = \{p_1, p_2, p_3\} $ and centers $ \mathcal{C} = \{c_1, c_2, c_3\} $ with unit quotas. The eligibility sets are:
\begin{itemize}
    \item \(E_{c_1}=\{p_1,p_2\}\)
    \item \(E_{c_2}=\{p_2,p_3\}\)
    \item  \(E_{c_3}=\{p_1\}\)
\end{itemize}

Under matching $ \mu $: $\mu(p_1) = c_1$, $\mu(p_2) = c_2$, $\mu(p_3) = \emptyset$.

The cycle $ \textbf{c} = (p_3, c_2, p_2, c_1, p_1, c_3, p_3) $ is unit eligibility improving since $p_3$ and $c_3$ are the unique unmatched parent and center, respectively.

\begin{figure}[h]
    \centering
    \begin{tikzpicture}[scale=1.2, every node/.style={font=\small}, thick,
        myedge/.style={->, >=stealth, shorten >=1pt}]

        \begin{scope}[xshift=-3cm]
            
            \node[draw, circle, inner sep=2pt] (p1_left) at (0,2) {$\mathbf{p_1}$};
            \node[draw, circle, inner sep=2pt] (p2_left) at (0,1) {$\mathbf{p_2}$};
            \node[draw, circle, inner sep=2pt, draw=gray!20, fill=gray!20] (p3_left) at (0,0) {$\mathbf{p_3}$}; 
            \node[draw, circle, inner sep=2pt] (c1_left) at (3,2) {$\mathbf{c_1}$};
            \node[draw, circle, inner sep=2pt] (c2_left) at (3,1) {$\mathbf{c_2}$};
            \node[draw, circle, inner sep=2pt, draw=gray!20, fill=gray!20] (c3_left) at (3,0) {$\mathbf{c_3}$}; 
            \draw[myedge, blue, thick] (c1_left) -- (p1_left);
            \draw[myedge, blue, thick] (c2_left) -- (p2_left);

            \draw[myedge, red, dashed] (p1_left) to[bend left=10] (c3_left);
            \draw[myedge, red, dashed] (p2_left) to[bend left=10] (c1_left);
            \draw[myedge, red, dashed] (p3_left) to[bend left=10] (c2_left);
            \draw[myedge, black, dashed] (c3_left) to[bend right=10] (p3_left);

            \node at (-0.5, 2.5) {$\mathbf{\mathcal{P}}$};
            \node at (3.5, 2.5) {$\mathbf{\mathcal{C}}$};
        \end{scope}

        \draw[->, thick, black] (0.7,1) -- (2.2,1) node[midway, above] {Apply Cycle $\textbf{c}$};

        \begin{scope}[xshift=3cm]
            
            \node[draw, circle, inner sep=2pt] (p1_right) at (0,2) {$\mathbf{p_1}$};
            \node[draw, circle, inner sep=2pt] (p2_right) at (0,1) {$\mathbf{p_2}$};
            \node[draw, circle, inner sep=2pt] (p3_right) at (0,0) {$\mathbf{p_3}$}; 
            \node[draw, circle, inner sep=2pt] (c1_right) at (3,2) {$\mathbf{c_1}$};
            \node[draw, circle, inner sep=2pt] (c2_right) at (3,1) {$\mathbf{c_2}$};
            \node[draw, circle, inner sep=2pt] (c3_right) at (3,0) {$\mathbf{c_3}$}; 
            \draw[myedge, blue, thick] (c1_right) -- (p2_right);
            \draw[myedge, blue, thick] (c2_right) -- (p3_right);
            \draw[myedge, blue, thick] (c3_right) -- (p1_right);

            \node at (-0.5, 2.5) {$\mathbf{\mathcal{P}}$};
            \node at (3.5, 2.5) {$\mathbf{\mathcal{C}}$};
        \end{scope}
    \end{tikzpicture}
    \caption{Left: Original associated directed graph $G_{\mu}$. Right: New matching after applying cycle $c$. Blue: matching edges ($c\to p$), Red dashed: eligibility edges ($p \to c$), Black dashed: unmatched edge \((c\to p)\).}
    \label{fig:corrected_cycle_example}
\end{figure}

Applying cycle $\textbf{c}$ yields: $\mathbf{c}(\mu)(p_1) = c_3$, $\mathbf{c}(\mu)(p_2) = c_1$, $\mathbf{c}(\mu)(p_3) = c_2$, demonstrating how cycles transform matchings while preserving eligibility.
\end{example}

Notice that, in the example above, after applying the cycle, the new matching is still eligible, and the total number of matches increases by one. This property holds by definition.

While the above definitions are standard in the matching literature, we now introduce \textbf{minimal cycles}, which are tailored to our two-tier environment and  characterize the non-domination frontier.

\begin{definition}
\label{def:minimal cycle}
    The \textbf{beneficiary loss} of applying a unit eligibility improving cycle $\mathbf{c}$ to undominated matching $\mu$ is $\Delta_{\mathbf{c}} = |B(\mu)| - |B(\mathbf{c}(\mu))| > 0$. A unit eligibility improving cycle is \textbf{minimal} if it achieves the smallest beneficiary loss among all unit eligibility improving cycles in $G_\mu$.
\end{definition}

Minimal cycles connect consecutive frontier points:

\begin{theorem}\label{theorem all cycles}
    Applying any minimal cycle $\mathbf{c}$ to any undominated matching $\mu_0$ yields an undominated matching $\mu_1 = \mathbf{c}(\mu_0)$.
\end{theorem}

\begin{proof}
    We prove by contradiction. Suppose $\mu_1$ is dominated. Then it must be dominated by some undominated matching $\mu_2$ with $|E(\mu_2)| \geq |E(\mu_1)|$. As $\mu_1$ is achieved by applying $\mathbf{c}$ to $\mu_0$, $|E(\mu_1)| = |E(\mu_0)| + 1$. Therefore, $|E(\mu_2)| \geq |E(\mu_1)| > |E(\mu_0)|$. 

We require the following result:

\begin{lemma}\label{disjoint cycle}
    For frontier points $f_1=(e_1,b_1),f_2=(e_1-k,b_2)$ where $k\in \mathbb{N}$, and any undominated matching $\hat{\mu}$ corresponding to $f_2$, there exist $k$ disjoint unit eligibility improving cycles in $G_{\hat{\mu}}$ that transform $\hat{\mu}$ to an undominated matching corresponding to \(f_1\). Each cycle has strictly positive beneficiary loss. 
\end{lemma}

We prove this lemma in Appendix~\ref{proof of  disjoint cycle}. By \autoref{disjoint cycle}, there exists $\mu_2^*$ with the same eligible and beneficiary match counts as $\mu_2$, obtained by applying disjoint unit eligibility improving cycles $\mathbf{c}_1, \ldots, \mathbf{c}_n$ to $\mu_0$, where
\(
n=|E(\mu_2)|-|E(\mu_0)|
\), each with strictly positive beneficiary loss. Thus $\mu^*_2$ also dominates $\mu_1$. The relationships are illustrated below:

    
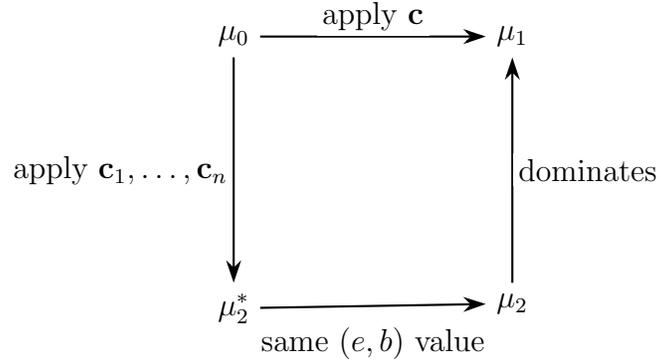
\begin{center}
\begin{tikzpicture}[
    node distance=3cm and 3cm,
    arrow/.style={thick, -{Stealth[length=3mm,width=2mm]}},
    label/.style={midway, fill=white, inner sep=2pt}
]

\node (mu0) {$\mu_0$};
\node[right=of mu0] (mu1) {$\mu_1$};
\node[below=of mu0] (mu2star) {$\mu_2^*$};
\node[below=of mu1] (mu2) {$\mu_2$};

\draw[arrow] (mu0) -- node[label, above] {apply $\textbf{c}$} (mu1);
\draw[arrow] (mu0) -- node[label, left] {apply $\textbf{c}_1,\dots,\textbf{c}_n$} (mu2star);
\draw[arrow] (mu2star) -- (mu2); 
\draw[arrow] (mu2) -- node[label, right] {dominates} (mu1);

\path (mu2star) -- (mu2) coordinate[pos=0.5] (m12);
\node[below=6pt of m12, fill=white, inner sep=2pt] {same $(e,b)$ value};

\end{tikzpicture}
\end{center}

    If $|E(\mu_2^*)| > |E(\mu_1)|$, then $n \geq 2$. This yields $|B(\mu_2^*)| = |B(\mu_0)| - \sum_{j=1}^n \Delta_{\mathbf{c}_j} < |B(\mu_0)| - \Delta_{\mathbf{c}_1} \leq |B(\mu_0)| - \Delta_{\mathbf{c}} = |B(\mu_1)|$, contradicting that $\mu_2^*$ dominates $\mu_1$.
    
    If $|E(\mu_2^*)| = |E(\mu_1)|$, then $n=1$ and $|B(\mu_2^*)| > |B(\mu_1)|$. But this means $\Delta_{\mathbf{c}_1} < \Delta_{\mathbf{c}}$, contradicting the minimality of $\mathbf{c}$. 
    
    Therefore, $\mu_1$ is undominated.
\end{proof}

After applying a minimal cycle to an undominated matching, the resulting matching has exactly one additional eligible match, corresponding to the next consecutive point on the frontier. Can we find the entire non-domination frontier by starting from an appropriate matching? The natural candidates are the frontier's endpoints.

A matching is \textbf{max-bene then max-elig} if it maximizes eligible matches among all matchings that maximize beneficiary matches. Let $\mathcal{M}_{BE}$ denote the set of such matchings:
    \[\mathcal{M}_{BE} = \underset{\mu \in \mathcal{M}}{\arg\max} |E(\mu)| \quad \text{where} \quad \mathcal{M} = \underset{\mu}{\arg\max} |B(\mu)|. \]
$\mathcal{M}_{BE}$ may contain multiple matchings. We fix an arbitrary $\mu_{BE} \in \mathcal{M}_{BE}$.

Analogously, a \textbf{max-elig then max-bene} matching $\mu_{EB}$ maximizes beneficiary matches among all matchings that maximize eligible matches. This corresponds to the solution in \cite{evren2023reserve}.

These matchings correspond to the frontier's endpoints. Since the frontier captures the complete trade-off between maximizing total matches and maximizing beneficiary matches, the endpoints represent the extremes of prioritizing each goal.

We say a frontier is \textbf{exhaustive in eligible matches} or exhibits \textbf{exhaustivity} if it contains points for every integer value of total matches from $|E(\mu_{BE})|$ to $|E(\mu_{EB})|$.

The following theorem establishes that repeatedly applying minimal cycles to $\mu_{BE}$ generates all frontier points:

\begin{theorem}\label{slope of the frontier}
    Starting from $\mu_{BE}$ and iteratively applying minimal cycles until no unit eligibility improving cycles remain generates matchings corresponding to all frontier points.
    The frontier of any instance is exhaustive in eligible matches.
    Moreover, beneficiary loss increases weakly along this sequence, so the frontier always exhibits concavity.
\end{theorem}
    
\begin{proof}
    By \autoref{disjoint cycle}, every undominated matching except $\mu_{EB}$ has a unit eligibility improving cycle. By \autoref{theorem all cycles}, applying a minimal cycle to an undominated matching yields another undominated matching with one additional eligible match. 
   Since no other undominated matching can exist between these two matchings, they are consecutive points on the frontier. Combining the above arguments, repeatedly applying minimal cycles generates all frontier points. Consequently, the frontier contains a point for every feasible integer value of total matches between $\lvert E(\mu_{BE})\rvert$ and $\lvert E(\mu_{EB})\rvert$.
   
   For the concavity of the frontier, consider the non-trivial case with three consecutive frontier points $(e_1,b_1),(e_1+1,b_2),(e_1+2,b_3)$ and any undominated matching $\mu_1$ corresponding to $(e_1,b_1)$. By \autoref{disjoint cycle}, there exist disjoint unit eligibility improving cycles $\mathbf{c}_1,\mathbf{c}_2$ in $G_{\mu_1}$ such that $b_3=b_1-\Delta _{\mathbf{c}_1}-\Delta _{\mathbf{c}_2}$. By \autoref{theorem all cycles}, there exists a minimal cycle $\mathbf{c}_3$ in $G_{\mu_1}$ with $b_2=b_1-\Delta _{\mathbf{c}_3}$ where $\Delta_{\mathbf{c}_3} \leq \min\{\Delta_{\mathbf{c}_1}, \Delta_{\mathbf{c}_2}\}$ (because it is minimal). 
    Therefore, $b_2-b_3=\Delta_{\mathbf{c}_1}+\Delta_{\mathbf{c}_2}-\Delta_{\mathbf{c}_3}\ge \Delta_{\mathbf{c}_3}=b_1-b_2$.
\end{proof}

\autoref{fig:frontier} illustrates the structure established above. The frontier consists of discrete points with integer coordinates. Connecting adjacent frontier points yields a continuous piecewise-linear curve, shown in blue. As total matches increase one unit at a time from $\mu_{BE}$ to $\mu_{EB}$, the corresponding loss in beneficiary matches grows weakly, so the slope of the piecewise curve is negative and decreasing.

\begin{figure}[h]
    \centering
    \begin{tikzpicture}[yscale=0.78, xscale=1.05, every node/.style={font=\small}]
        \coordinate (P0) at (2,8.2);  
        \coordinate (P1) at (3,7.6);
        \coordinate (P2) at (4,6.6);
        \coordinate (P3) at (5,5.2);
        \coordinate (P4) at (6,3.8);
        \coordinate (P5) at (7,1.4);  

        \foreach \p in {P0,P1,P2,P3,P4,P5} {
            \draw[dashed, gray!60] (\p) -- (\p |- 0,0);
            \draw[dashed, gray!60] (\p) -- (0,0 |- \p);
        }

        \node[left] at (0,8.2) {$b_0$};
        \node[left] at (0,7.6) {$b_1$};
        \node[left] at (0,6.6) {$b_2$};
        \node[left] at (0,5.2) {$b_3$};
        \node[left] at (0,3.8) {$b_4$};
        \node[left] at (0,1.4) {$b_5$};

        \draw[decorate, decoration={brace, amplitude=5pt}, line width=1pt]
            (-0.9,7.6) -- (-0.9,8.2) node[midway, left=4pt] {$\Delta b_1$};
        \draw[decorate, decoration={brace, amplitude=5pt}, line width=1pt]
            (-0.9,6.6) -- (-0.9,7.6) node[midway, left=4pt] {$\Delta b_2$};
        \draw[decorate, decoration={brace, amplitude=5pt}, line width=1pt]
            (-0.9,5.2) -- (-0.9,6.6) node[midway, left=4pt] {$\Delta b_3$};
        \draw[decorate, decoration={brace, amplitude=5pt}, line width=1pt]
            (-0.9,3.8) -- (-0.9,5.2) node[midway, left=4pt] {$\Delta b_4$};
        \draw[decorate, decoration={brace, amplitude=5pt}, line width=1pt]
            (-0.9,1.4) -- (-0.9,3.8) node[midway, left=4pt] {$\Delta b_5$};

        \node[below] at (2,0) {$e_0$};
        \node[below] at (3,0) {$e_0{+}1$};
        \node[below] at (4,0) {$e_0{+}2$};
        \node[below] at (5,0) {$e_0{+}3$};
        \node[below] at (6,0) {$e_0{+}4$};
        \node[below] at (7,0) {$e_0{+}5$};

        \draw[thick, blue] (P0) -- (P1) -- (P2) -- (P3) -- (P4) -- (P5);
        \foreach \p in {P0,P1,P2,P3,P4,P5}
            \fill[blue] (\p) circle (2.4pt);

        \node[above right] at (P0) {$\mu_{BE}$};
        \node[below left]  at (P5) {$\mu_{EB}$};

        \draw[->, thick] (0,0) -- (8.2,0) node[right] {$e$ (total matches)};
        \draw[->, thick] (0,0) -- (0,9.0) node[above] {$b$ (beneficiary matches)};

        \fill[white] (0.85,-0.08) -- (1.15,-0.08) -- (1.35,0.08) -- (1.05,0.08) -- cycle;
        \draw[thick] (0.8,-0.12) -- (1.0,0.12);
        \draw[thick] (1.0,-0.12) -- (1.2,0.12);
    \end{tikzpicture}
    \caption{The non-domination frontier as a concave, piecewise-linear curve. The break on the horizontal axis indicates that the initial portion of the horizontal axis is omitted (in particular, every frontier point satisfies $b \le e$). Total matches increase in unit steps along the horizontal axis (from $e_0$ at $\mu_{BE}$ to $e_0+5$ at $\mu_{EB}$); the dashed lines map each frontier point to its beneficiary level $b_0,\dots,b_5$ on the vertical axis. The braces mark the per-step beneficiary loss $\Delta b_k = b_{k-1}-b_k$, which is weakly increasing: here $\Delta b_1 < \Delta b_2 < \Delta b_3 = \Delta b_4 < \Delta b_5$. Equivalently, the frontier exhibits concavity.}
    \label{fig:frontier}
\end{figure}

The following proposition quantifies the worst-case resource sacrifice when prioritizing beneficiary matches. 

\begin{proposition}\label{Upper bond}
For any instance $(\mathcal{C},\mathcal{P},\textbf{q},\mathcal{E},\mathcal{B})$:
\[
\frac{|E(\mu_{EB})|-|E(\mu_{BE})|}{|\mathcal{P}|}
\le
\frac{|E(\mu_{EB})|-|E(\mu_{BE})|}
{\min\left(|\mathcal{P}|,\sum_{c\in\mathcal{C}}q_c\right)}
\le
\frac{|E(\mu_{EB})|-|E(\mu_{BE})|}{|E(\mu_{EB})|}
\le
\frac{1}{2}.
\]
This bound is tight: \autoref{conflict} achieves
\(\frac{|E(\mu_{EB})|-|E(\mu_{BE})|}{|\mathcal{P}|}
=
\frac{1}{2}.\)

\end{proposition}

\begin{proof}
Let \(K=|E(\mu_{EB})|-|E(\mu_{BE})|\). By
\autoref{disjoint cycle}, there exist \(K\) disjoint unit eligibility
improving cycles transforming \(\mu_{BE}\) to some max-elig then
max-bene matching \(\mu_1\).

Each cycle contains at least two parents---a single-parent cycle would
allow increasing eligible matches without changing other matches,
contradicting that \(\mu_{BE}\) is undominated. Therefore, \(\mu_1\)
matches at least \(2K\) distinct parents, yielding
\[
\frac{|E(\mu_{EB})|-|E(\mu_{BE})|}{|E(\mu_{EB})|}
\le
\frac{1}{2}.
\]

The result follows since
\(|\mathcal{P}|
\ge
\min\left(|\mathcal{P}|,\sum_{c\in\mathcal{C}}q_c\right)
\ge
|E(\mu_{EB})|.\)
\end{proof}

This bound has profound policy implications: among undominated matchings, at most half the matched seats can be sacrificed under  {any} policy prioritizing beneficiary matches---whether through quotas, beneficiary-share guarantees, or weighted objectives combining total and beneficiary matches.

\subsection{Repeated Hungarian Algorithm}
While directly identifying minimal cycles to calculate the frontier becomes computationally challenging as the number of parents and seats grows, the characterization enabled by these cycles allows for a novel approach to frontier calculation,\footnote{Alternatively, consider it as a linear programming problem: loop over the number of total matches $e$, maximize the number of beneficiary matches subject to the number of total matches being greater than or equal to the current $e$.} through repeated application of the Hungarian algorithm.

From \autoref{slope of the frontier}, the non-domination frontier has decreasing slope and hence is concave. To find all frontier points, we first identify kinks where the slope changes, then exploit the exhaustivity property to fill intermediate points by linear interpolation.\footnote{That is, since the frontier is exhaustive in eligible matches, we fill in intermediate points at every unit interval in eligible matches by extending the same slope between them.}

To find all the kinks, we employ the Hungarian algorithm, typically used to compute a maximum-weight matching in a bipartite graph \citep{kuhn1955hungarian}.\footnote{While the classical Hungarian algorithm formulation assumes equal-sized partitions, the standard solution for unbalanced cases involves adding dummy vertices connected by zero-weight edges to create an equivalent balanced bipartite graph, thus enabling the algorithm's application.} Our bipartite graph has parents $\mathcal{P}$ on one side, centers $\mathcal{C}$ on the other, with edges connecting eligible parent--center pairs regardless of beneficiary status. Edge weights determine which frontier kink is identified---the algorithm finds points where supporting lines of different slopes meet the frontier.\footnote{A supporting line touches the frontier at exactly one point.}

The complete procedure is presented in \autoref{alg:RHA}.

\begin{algorithm}[!t]
\caption{Repeated Hungarian Algorithm (RHA)}\label{alg:RHA}
\begin{algorithmic}[1]
\State Initialize frontier $\gets \emptyset$
\State Set $n \gets \max(|\mathcal{P}|,\sum_{c\in\mathcal{C}}q_c)$
\State Set $m \gets 0$ \textit{ (Counter of kinks)}

\For{$k \gets 1$ to $n$}
    \State Assign weights:
    \State \hspace{1em} Eligible but not beneficiary:
    $w_{\text{elig}} \gets 1$
    \State \hspace{1em} Beneficiary:
    $w_{\text{bene}} \gets 1+\frac{1}{k}+\frac{1}{n^2}$
    \State Run the Hungarian algorithm and let $\mu$ be a matching that maximizes
    $w_{\text{elig}}\cdot (|E(\mu)|-|B(\mu)|)
    +w_{\text{bene}}\cdot |B(\mu)|$
    \State Compute $e \gets |E(\mu)|$
    \State Compute $b \gets |B(\mu)|$

    \If{$m=0$}
        \State $m\gets m+1$
        \State $(e_m,b_m)\gets(e,b)$
        \State frontier $\gets$ frontier $\cup\{(e,b)\}$
    \ElsIf{$(e,b)\neq(e_m,b_m)$}
        \State $m\gets m+1$
        \State $(e_m,b_m)\gets(e,b)$
        \State frontier $\gets$ frontier $\cup\{(e,b)\}$
    \EndIf
\EndFor

\For{$k \gets 1$ to $m-1$}
    \State \textit{(Insert intermediate points between kink
    $(e_k,b_k)$ and kink $(e_{k+1},b_{k+1})$)}
    \For{$j\gets 1$ \text{ to } $e_{k+1}-e_k-1$}
        \State frontier $\gets$ frontier $\cup
        \left\{
        \left(
        e_k+j,\,
        b_k-j\cdot\frac{b_k-b_{k+1}}{e_{k+1}-e_k}
        \right)
        \right\}$
    \EndFor
\EndFor
\end{algorithmic}
\end{algorithm}

To establish correctness of this algorithm, note that by \autoref{slope of the frontier}, the frontier $F$ for any instance takes the form:
\[ F = \left\{ (e, b), (e - 1, b + \Delta b_1), \ldots, (e - K, b + \sum_{i=1}^K \Delta b_i) \right\} \]

where $K = |E(\mu_{EB})| - |E(\mu_{BE})|$. This frontier contains $K+1$ points. The starting point $(e,b)$ corresponds to $\mu_{EB}$, while the endpoint $(e-K, b+\sum_{i=1}^K \Delta b_i)$ corresponds to $\mu_{BE}$. The beneficiary differences satisfy $\Delta b_1 \geq \Delta b_2 \geq \cdots \geq \Delta b_K \geq 1$.

Point $j$ is a kink when $\Delta b_j \neq \Delta b_{j+1}$. At kink $(e-j, b+\sum_{i=1}^j \Delta b_i)$, the marginal trade-off changes from $\Delta b_j$ to $\Delta b_{j+1}$ beneficiary matches per eligible match. This point lies on the supporting line with slope in $(-\Delta b_{j}, -\Delta b_{j+1})$, which corresponds to beneficiary weight $w_\text{bene} = 1 + \frac{1}{\Delta b_j} + \frac{1}{n^2}$ in the algorithm.

\begin{lemma}\label{slope change points}
The Repeated Hungarian Algorithm (RHA) satisfies:
\begin{enumerate}
    \item In the first loop, the first and last iterations compute $\mu_{BE}$ and $\mu_{EB}$ respectively.
    \item For any interior frontier point $f_j = (e-j, b+\sum_{i=1}^j \Delta b_i)$ (excluding endpoints) with $\Delta b_j > \Delta b_{j+1}$, the RHA computes this point in iteration $\Delta b_j$ of the first loop.
\end{enumerate}
\end{lemma}

We prove this lemma in Appendix~\ref{proof of  slope change points}. We now establish the algorithm's properties.

\begin{theorem}\label{Property of RHA}
The Repeated Hungarian Algorithm (RHA) satisfies:
\begin{enumerate}
\item \textbf{Completeness}: RHA generates the entire non-domination frontier.
    
    \item \textbf{Computational Complexity}: RHA runs in \( O\left(n^4\right) \) time, where $ n= \max(|\mathcal{P}|,\sum _{c\in \mathcal{C} }q_c)$.
\end{enumerate}
\end{theorem}

\begin{proof}

For completeness, \autoref{slope change points} establishes that the first loop identifies all kinks (points where slope changes) and both endpoints. Any remaining frontier points lie on constant-slope segments and are filled by linear interpolation in the second loop.

The complexity bound follows from the first loop, which runs at most $n$ iterations (since $\Delta b_i \geq 1$ and there are at most $n$ possible values), with each iteration executing the Hungarian algorithm in $O(n^3)$ time \citep{kuhn1955hungarian}. The second loop performs only linear interpolation in $O(n)$ time. Thus, the total complexity is $O(n^4)$. 
\end{proof}

\begin{remark}\label{Concavity and RHA}
Frontier concavity is essential for our approach. Suppose the frontier is not concave (but is still exhaustive in eligibility), that is, there exist three hypothetical points \(f_1=(e, b)\), \(f_2=(e-1, b+\Delta b_1)\), and \(f_3=(e-2, b+\Delta b_1+\Delta b_2)\) with \(\Delta b_1 < \Delta b_2\). To identify $f_2$ in $RHA$, there exists beneficiary weight \(w\in \mathbb{R}_{++}\) at which $f_2$ exceeds both $f_1$ and $f_3$ in weighted value, requiring 
\[w(\Delta b_1)-1>0\quad\text{and}\quad 1-w(\Delta b_2)>0.\]
These inequalities cannot hold simultaneously when $\Delta b_1 < \Delta b_2$. Thus, without concavity, RHA cannot identify all the kinks and fails to generate the complete frontier.
\end{remark}

\section{Compatibility with Design Objectives}

This section examines how the non-domination frontier interacts with key allocation criteria that arise in practice. We first address the tension between achieving beneficiary-share guarantees and maintaining undominated allocations. We then study whether the framework can accommodate priority orderings and whether parent-side eligibility information can be elicited in a strategy-proof manner. We next examine path-independence, a foundational property for stable market design. Finally, we introduce parents' strict preferences over centers and study Pareto efficiency and fairness.
\subsection{Beneficiary-Share Guarantee and Non-Domination}\label{sec:beneguar}

We now augment the model with a policy instrument, the beneficiary-share guarantee. 
For any non-empty matching $\mu$, define its \textbf{beneficiary-share} to be
\[
\beta(\mu) =\frac{|B(\mu)|}{|E(\mu)|},
\]
the fraction of total matches that are beneficiary matches. A \textbf{beneficiary-share guarantee} is a number $\beta^* \in [0,1]$ chosen by the policymaker, specifying the minimum share of total matches that must go to beneficiaries; together with an instance it forms a \textbf{problem} $(\mathcal{C},\mathcal{P},\textbf{q},\mathcal{E}, \mathcal{B}, \beta^*)$.  When a non-empty matching $\mu$ satisfies $\beta(\mu)\ge \beta^*$, we say it \textbf{respects the beneficiary-share guarantee}.

A mechanism $\varphi$ \textbf{respects the beneficiary-share guarantee} if for any problem with at least one non-empty beneficiary parent set, the mechanism chooses a matching that respects the beneficiary-share guarantee.\footnote{For any problem with at least one non-empty beneficiary parent set, a matching respecting the beneficiary-share guarantee exists: simply match only within this beneficiary parent set, achieving $\beta(\mu) = 1 \geq \beta^*$. Thus, mechanisms respecting the beneficiary-share guarantee exist.}

While such mechanisms exist, the beneficiary-share guarantee fundamentally conflicts with non-domination, as the following example demonstrates.

\begin{example}\label{beta threshold}
Consider a problem with parents $\mathcal{P}=\{p_1,p_2\}$ and centers $\mathcal{C}=\{c_1,c_2\}$ with $q_{c_1} = q_{c_2} = 1$, and a beneficiary-share guarantee $\beta^* =0.7$. Parent $p_1$ is a beneficiary of $c_1$ but not eligible for $c_2$, while parent $p_2$ is eligible for $c_2$ but not a beneficiary. 

To respect the beneficiary-share guarantee, any mechanism must choose a matching $\mu$ with $\mu(p_1)=c_1$ and $\mu(p_2)=\emptyset$, yielding $\beta(\mu)=1$. The alternative matching $\mu'$ with $\mu'(p_1) = c_1$ and $\mu'(p_2) = c_2$ violates the guarantee since $\beta(\mu') = 0.5 < 0.7$.

However, $\mu'$ dominates $\mu$ by achieving $(|E(\mu')|, |B(\mu')|) = (2, 1)$ versus $(|E(\mu)|,$ $|B(\mu)|) = (1, 1)$. In this case, the beneficiary-share guarantee forces selection of a dominated matching.
\end{example}

\autoref{beta threshold} demonstrates a critical policy limitation: mandating too high a beneficiary-share guarantee can force selection of a matching dominated by $\mu_{BE}$. This domination is unjustifiable---while the beneficiary-share guarantee aims to increase beneficiary matches, the resulting matching achieves no more beneficiary matches than $\mu_{BE}$ while achieving strictly fewer total matches.

With this consideration, when non-domination conflicts with the beneficiary-share guarantee, we prioritize non-domination. We formalize this through the following concept: given a problem, a matching $\mu$ \textbf{respects the beneficiary-share guarantee approximately on the frontier} if:
\begin{itemize}
\item When $\beta^*<\beta(\mu_{BE})$: $\mu$ achieves the beneficiary-share closest to $\beta^*$ among all undominated matchings satisfying $\beta(\mu) \geq \beta^*$.
\item When $\beta^*\geq \beta(\mu_{BE})$: $\mu$ is a max-bene then max-elig matching.
\end{itemize}

A mechanism \textbf{respects the beneficiary-share guarantee approximately on the frontier} if it always selects such a matching.\footnote{By definition, $\beta(\mu_{BE}) = 0$ when all beneficiary parent sets are empty.}\textsuperscript{,}\footnote{Our solution concept satisfies Pareto optimality under dichotomous matched-versus-unmatched preferences. A matching $\mu$ is Pareto dominated by $\mu'$ if: (i) every participant matched in $\mu$ remains matched in $\mu'$ (i.e., $\mu(p) \neq \emptyset \implies \mu'(p) \neq \emptyset$), and (ii) at least one previously unmatched participant becomes matched (i.e., $\exists p' \in \mathcal{P}$ such that $\mu(p') = \emptyset$ but $\mu'(p') \neq \emptyset$). Since the selected approximately compliant frontier matching maximizes total matches subject to the beneficiary-share guarantee, it cannot be Pareto dominated by any other matching that respects the same guarantee---any additional match would violate the beneficiary-share constraint.} Such mechanisms always exist by construction.

The following result further justifies this solution concept: when we could choose a dominated matching $\mu$ achieving exactly $\beta^*$, selecting instead a matching that respects the beneficiary-share guarantee approximately on the frontier yields a matching that dominates $\mu$.

\begin{proposition}\label{justification of approximation}
    For any problem $(\mathcal{C},\mathcal{P},\textbf{q},\mathcal{E}, \mathcal{B}, \beta^*)$, let \(\mu^*\) be a matching that respects the beneficiary-share guarantee approximately on the frontier. If \(\beta(\mu^*)\neq \beta^*\), then \(\mu^*\) dominates any matching with beneficiary-share \(\beta^*\).
\end{proposition}

We prove this proposition in Appendix~\ref{proof of justification of approximation}.

\subsection{Respecting Priority and Strategy-Proofness}

In many real-world allocation systems, reserves operate alongside priority orderings of parents. This subsection extends our model to incorporate such priorities. Following \citet{pathak2024}, we allow center-specific priorities while requiring beneficiary parents to rank above non-beneficiary parents within each center.

Formally, a list of strict total orders on $\mathcal{P}$, denoted $\pi=(\pi_c)_{c\in \mathcal{C}}$, is a \textbf{priority order} if for any center $c$:
\begin{itemize}
\item For any $(p,j)\in B_c\times(E_c\setminus B_c)$: $p \mathrel{\pi_c} j$ (beneficiaries rank above non-beneficiaries), where $p\pi_c j$ means that parent $p$ has higher priority than parent $j$ at center $c$
\item For any $(j,k)\in E_c\times (\mathcal{P}\setminus E_c)$: $j \mathrel{\pi_c} k$ (eligible parents rank above ineligible parents)
\end{itemize}

We call the vector $(\mathcal{C},\mathcal{P},\textbf{q},\mathcal{E}, \mathcal{B},\beta^*,\pi)$ a \textbf{problem with order}. With slight abuse of notation, we retain our terminology for mechanisms, matchings, and respecting the beneficiary-share guarantee approximately on the frontier when applied to problems with order.

Given a problem with order, a matching $\mu$ \textbf{respects priority} if for every center $c$ and parents $p, p'\in \mathcal{P}$:

\[
\mu(p)=c \text{ and }\mu(p')=\emptyset \implies p\pi_cp'.
\]
This condition ensures no unassigned parent has higher priority than an assigned parent for the same center, preventing justified envy. This definition aligns with priority-respecting matching in \citet{pathak2024}. A mechanism $\varphi$ \textbf{respects priority} if for any problem with order $(\mathcal{C},\mathcal{P},\textbf{q},\mathcal{E}, \mathcal{B},\beta ^*,\pi)$, the mechanism's output respects priority.

The following result establishes that respecting priority is compatible with respecting the beneficiary-share guarantee approximately on the frontier.

\begin{proposition}\label{Respecting priority}
    There exists a mechanism that respects priority and respects the beneficiary-share guarantee approximately on the frontier.
\end{proposition}

We prove this proposition in Appendix~\ref{proof of Respecting priority}.

The preceding analysis takes the pair-specific eligibility relation as given. We now ask whether its family-side component can instead be privately known and truthfully elicited while preserving the properties established above.

For each center $c$, let $\bar E_c$ denote the set of parents who satisfy the center-side admissibility requirements. For each parent $p_i$, let $T_i\subseteq\mathcal{C}$ denote the set of day care centers that the parent is willing to accept. The set $T_i$ is privately known to the parent, while center-side admissibility, beneficiary status, quotas, and priority orders remain fixed. Each parent reports a set $A_i\subseteq\mathcal{C}$. Given a report profile $A=(A_i)_{p_i\in\mathcal P}$, define
\[
E_c(A)=\{p_i\in\bar E_c:c\in A_i\},
\qquad
B_c(A)=B_c\cap E_c(A).
\]
Let $\mathcal E(A)=(E_c(A))_{c\in\mathcal C}$ and
$\mathcal B(A)=(B_c(A))_{c\in\mathcal C}$.

 A matching under report profile $A$ respects priority if no unmatched parent who is eligible for a center under $A$ has higher priority there than a parent assigned to that center.

For any report profile $A$, let $\varphi(A)$ denote the matching
selected by the mechanism under $\mathcal E(A)$ and $\mathcal B(A)$,
holding all other primitives fixed. We say that $\varphi$ is
\textbf{strategy-proof} if, for every parent $p_i$, every true
acceptable set $T_i\subseteq\mathcal C$, every report profile $A_{-i}$
of the other parents, and every alternative report
$A_i'\subseteq\mathcal C$,
\[
\varphi(T_i,A_{-i})(p_i)=\emptyset
\quad\Longrightarrow\quad
\varphi(A_i',A_{-i})(p_i)\notin T_i.
\]
That is, no parent who would remain unmatched under truthful reporting
can obtain an assignment in her true acceptable set by misreporting.
Thus, the mechanism elicits only the family-side component of
eligibility rather than a ranking over day care centers. The following
result shows that this information can be elicited without sacrificing
either priority or our frontier-based beneficiary-share criterion.

\begin{proposition}\label{Strategy-proof}
There exists a mechanism that respects priority, respects the beneficiary-share guarantee approximately on the non-domination frontier, and is strategy-proof.
\end{proposition}

The proof is provided in Appendix~\ref{proof of  Strategy-proof}. Thus, both priority and truthful elicitation of family-side eligibility information are compatible with our frontier-based solution. 

\subsection{Path-independence}
This subsection analyzes path-independence, a fundamental property of choice rules. We first define the choice rule induced by a mechanism.

Given a problem $(\mathcal{C},\mathcal{P},\textbf{q},\mathcal{E}, \mathcal{B},\beta ^*)$, the \textbf{choice rule }\(C\) induced by a mechanism $\varphi$ is a function from \(2^{\mathcal{P}}\) to \(2^\mathcal{P}\) such that for any \(X \in 2^\mathcal{P}\):   \[C(X) =\{p\in X:  \varphi(\mathcal{C},X,\textbf{q},\mathcal{E}, \mathcal{B},\beta ^*)(p)\neq \emptyset\}.\] 
That is, $C(X)$ consists of all parents in $X$ who receive assignments when the mechanism is applied to parent set $X$.

Given a problem, a choice rule \(C\) is \textbf{path-independent} if for all \(X, X' \in 2^\mathcal{P}\):
\[C(X \cup X') = C(C(X) \cup X').\] 
This property ensures that choosing from a union can be done sequentially: first choosing from $X$, then combining those chosen with $X'$ and choosing again yields the same result as choosing directly from $X \cup X'$. Path-independence plays a fundamental role in market design, particularly for choice rule-based deferred acceptance mechanisms \citep{dougan2025market}. A mechanism is \textbf{path-independent} if the choice rule it induces is path-independent for every problem.

\begin{theorem}\label{thrm PI imposible}
    There does not exist a mechanism that is path-independent and respects the beneficiary-share guarantee approximately on the frontier.
\end{theorem}

We prove this theorem in Appendix~\ref{proof of thrm PI imposible}.

\subsection{Efficiency and Fairness}\label{preferences and efficiency}

In this subsection, we introduce parents' preferences over centers and show that frontier allocations need not be Pareto efficient or fair.

Endow each parent $p\in\mathcal P$ with a strict preference relation
$\succ_p$ over $\mathcal C\cup\{\emptyset\}$. We assume that every parent
strictly prefers any center for which she is eligible to remaining
unmatched, and strictly prefers remaining unmatched to any center for
which she is ineligible. For each $c\in\mathcal C$, let $R_p(c)$ denote
the rank of center $c$ in the restriction of $\succ_p$ to $\mathcal C$,
with $R_p(c)=1$ for $p$'s most-preferred center. The relative ranking of
ineligible centers is immaterial because we only consider eligible
matchings. In the preference lists below, we therefore omit centers where the corresponding parents are ineligible.

First, the following example demonstrates that non-domination may admit no Pareto efficient matching.

\begin{example}\label{nd not pareto}
Let $\mathcal{P}=\{p_1,p_2\}$ and $\mathcal{C}=\{c_1,c_2\}$ with unit quotas. Both parents are eligible for both centers. The beneficiary parent sets are $B_{c_1}=\{p_1\}$ and $B_{c_2}=\emptyset$, so $p_1$ is a beneficiary of $c_1$ alone. Preferences are
\[
p_1:\; c_2\succ_{p_1} c_1\succ_{p_1}\emptyset,
\qquad
p_2:\; c_1\succ_{p_2}c_2\succ_{p_2}\emptyset.
\]
Observe that the unique undominated matching \(\mu=\{(p_1,c_1),(p_2,c_2)\}\) is Pareto dominated by \(\mu'=\{(p_1,c_2),(p_2,c_1)\}\).
\end{example}

Second, the following example shows there may be no fair matching on the frontier. Suppose each center $c$ ranks the parents through a strict priority order $\pi_c$, a linear order over $\mathcal{P}$ in which every beneficiary of $c$ is ranked above every non-beneficiary of $c$. Parent $p$ has \textbf{justified envy} at $\mu$ if there is a center $c$ with $c\succ_p\mu(p)$ at which $p$ has higher $\pi_c$-priority than some parent in $\mu(c)$, or at which a seat is vacant; a matching is \textbf{fair} if no parent has justified envy. 

\begin{example}\label{nd not fair}
Let $\mathcal{P}=\{p_A,p_B\}$ and $\mathcal{C}=\{c_1,c_2\}$ with unit quotas. Parent $p_A$ is eligible for both centers and is a beneficiary of both, $p_A\in B_{c_1}\cap B_{c_2}$; parent $p_B$ is eligible only for $c_2$ and is a beneficiary of $c_2$, $p_B\in B_{c_2}$. At center $c_2$, let $p_A\pi_{c_2}p_B$. Preferences are
\[
p_A:\; c_2\succ_{p_A} c_1\succ_{p_A}\emptyset,
\qquad
p_B:\; c_2\succ_{p_B}\emptyset.
\]
The unique matching attaining two beneficiary matches assigns
$\mu(p_A)=c_1$ and $\mu(p_B)=c_2$, yielding
$|E(\mu)|=|B(\mu)|=2$. However, under it $p_A$ has justified envy toward $c_2$: parent $p_A$ prefers $c_2$ to her assigned $c_1$ and has higher $\pi_{c_2}$-priority than the incumbent $p_B$. Hence non-domination and fairness cannot in general be satisfied together.
\end{example}

In practice, it is not uncommon to sacrifice fairness and efficiency to achieve other important objectives. The Union Public Service Commission (of India) abandoned the (only) fair and efficient serial dictatorship mechanism to address regional homophily in the Indian Administrative Service \citep{bo2026visiblyfairmechanisms}. 

However, the designer can still deliver a constrained efficient matching. A matching is \textbf{constrained efficient at $(e,b)$} if it corresponds to some $(e,b)$ on the non-domination frontier and is not Pareto dominated by any other matching corresponding to $(e,b)$. As we will show, such a matching can be obtained as the solution to a linear program. For each parent $p$ and center $c$ introduce a variable $x_{pc}\in[0,1]$, interpreted as matching $p$ to $c$, and let $M$ be a constant larger than $\sum_{p,c}R_p(c)$. Consider the linear program
\begin{align}
\max_{x}\quad & \sum_{p\in\mathcal{P}}\sum_{c\in\mathcal{C}}\bigl(\,\mathbf{1}\{p\in B_c\}\cdot M - R_p(c)\,\bigr)\,x_{pc} \label{lp:ce}\\
\text{s.t.}\quad
& \sum_{c\in\mathcal{C}} x_{pc}\le 1 && \forall\, p\in\mathcal{P},\notag\\
& \sum_{p\in\mathcal{P}} x_{pc}\le q_c && \forall\, c\in\mathcal{C},\notag\\
& \sum_{p\in\mathcal{P}}\sum_{c\in\mathcal{C}} x_{pc}=e,\notag\\
& x_{pc}=0 && \text{if } p\notin E_c,\notag\\
& x_{pc}\ge 0 && \forall\, p\in\mathcal{P},\,c\in\mathcal{C}.\notag
\end{align}
The first two families of constraints are the parent- and center-capacity constraints; the equality constraint fixes the total number of matches at the target level $e$; and the eligibility restriction forbids ineligible pairs. The large coefficient $M$ first enforces the maximal beneficiary count $b$ associated with $e$ on the frontier, after which the rank term $-R_p(c)$ selects, among matchings attaining $(e,b)$, one of minimal aggregate preference rank. The correctness and efficiency of linear program \eqref{lp:ce} are established by the following result, which we prove in Appendix \ref{proof of  prop:ce lp}.

\begin{proposition}\label{prop:ce lp}
Every integral optimal solution of linear program \eqref{lp:ce}
induces a matching that is constrained efficient at $(e,b)$, and such
a solution can be computed in polynomial time.
\end{proposition}

\begin{remark}
The tensions in Examples 4 and 5 arise specifically between the designer's objectives and parents' individual welfare. If the designer is instead counted as an agent who weakly prefers any matching with weakly more total and beneficiary matches, then every constrained efficient matching is Pareto efficient in this broader sense: a matching that leaves the designer weakly better off must attain the same frontier point $(e,b)$ by non-domination, so the strict improvement would have to come from a parent, contradicting constrained efficiency. The converse does not hold---the matching $\mu'$ in Example 4 is efficient in this broader sense despite lying strictly below the frontier, since reaching the frontier requires hurting parent $p_1$. 
\end{remark}

Lastly, if preferences are not drawn from the universal domain, one can obtain globally efficient matchings. The conflict with Pareto efficiency in \autoref{nd not pareto} stems from a parent preferring a center for which she is not a beneficiary. In the day care setting, parents may instead prefer centers closer to home, which are exactly the centers for which they are beneficiaries. Such preferences naturally arise in many settings, for instance, parking slot assignment.\footnote{Thanks to Vikram Manjunath for mentioning this.} Under such a restriction on the preference domain, at every point on the frontier there exists a Pareto efficient matching. 

Formally, we say preferences are \textbf{beneficiary-aligned} if for every parent $p$ and centers $c,c'$, whenever $p\in B_c$ and $p\notin B_{c'}$ we have $c\succ_p c'$. That is, every center for which $p$ is a beneficiary is preferred to every center for which $p$ is not.\footnote{Equivalently, ranking centers by ``$p$ is a beneficiary'' versus ``$p$ is only eligible'' and then arbitrarily within each class.} Under this restriction, efficiency is recovered.

\begin{proposition}\label{aligned efficiency}
If preferences are \textbf{beneficiary-aligned}, then for every point $(e,b)\in F$ there exists a matching corresponding to $(e,b)$ that is Pareto efficient.
\end{proposition}
We prove \autoref{aligned efficiency} in Appendix \ref{proof of  aligned efficiency}.

However, alignment does not restore fairness. In fact, the preferences in \autoref{nd not fair} are themselves beneficiary-aligned---each of $p_A$ and $p_B$ ranks her beneficiary centers above any non-beneficiary center---yet the frontier matching there still admits justified envy. The restriction therefore reconciles non-domination with Pareto efficiency without resolving the conflict with fairness.

\section{Conclusion}

We characterize the non-domination frontier between total matches and beneficiary matches through the novel concept of minimal cycles. Our analysis reveals that the frontier exhibits a concave structure with increasing marginal costs---each additional eligible match requires progressively larger sacrifices in beneficiary matches. We establish that the relative efficiency loss from beneficiary-share requirements is bounded by $\frac{1}{2}$, and this bound is tight. The Repeated Hungarian Algorithm we develop generates all frontier points in polynomial time ($O(n^4)$), providing practitioners with a computationally tractable tool. Our framework demonstrates that respecting beneficiary-share guarantees approximately on the frontier is compatible with priority considerations but fundamentally incompatible with path-independent choice rules. 

While our analysis focuses on settings with a single layer of beneficiary status, the minimal cycle approach may extend to richer environments. Multiple overlapping beneficiary requirements---such as simultaneous quotas for different demographic groups in school allocation---present interesting directions for future work. Our characterization moves beyond traditional lexicographic approaches, offering guidance for balancing competing objectives. The minimal cycle construction we introduce provides a new analytical tool for matching theory that may prove valuable in broader reserve system design.

\newpage

\textbf{Declaration of generative AI and AI-assisted technologies in the manuscript preparation process:}
We used LLMs developed by Anthropic, OpenAI, and Kimi to polish the paper’s writing. We reviewed and edited the output as needed and take full responsibility for the content of the published article. 

\bibliographystyle{ecca}

\bibliography{references}
\newpage
\appendix
\section{Appendix}\label{appendix}
\renewcommand{\thelemma}{A\arabic{lemma}}
\renewcommand{\theHlemma}{A.\arabic{lemma}}
\setcounter{lemma}{0}

\renewcommand{\theproposition}{A\arabic{proposition}}
\renewcommand{\theHproposition}{A.\arabic{proposition}}
\setcounter{proposition}{0}

\subsection{\autoref{shorcoming of max in max}}

While it may seem natural to first maximize the total number of matches and then maximize beneficiary matches among those solutions, this lexicographic approach can lead to poor outcomes for beneficiaries. We demonstrate that prioritizing total matches can result in an arbitrarily large loss of beneficiary matches compared to directly maximizing beneficiary matches.

\begin{example}
\label{shorcoming of max in max}
Consider the following market instance with parameter $K \in \mathbb{N}$:

\begin{itemize}
    \item  $\mathcal{P} = \{p_1, p_2, \ldots, p_{K+2}\}$.
    \item  $\mathcal{C} = \{c_1, c_2, \ldots, c_{K+2}\}$, each with a unit quota.
\end{itemize}

The beneficiary and eligibility sets are defined as follows:
\begin{align*}
    B_{c_n} &= \{p_n\} \quad \text{for } n = 1, 2, \ldots, K+1, \quad &B_{c_{K+2}} &= \emptyset, \\
    E_{c_n} &= \{p_n, p_{n-1}\} \quad \text{for } n = 2, \ldots, K+1, \quad &E_{c_1} &= \{p_1, p_{K+2}\}, \quad E_{c_{K+2}} = \{p_{K+1}\}.
\end{align*}

In this construction:
\begin{itemize}
    \item A \textbf{max-bene matching} $\mu_1$ can assign each parent $p_n$ ($1 \leq n \leq K+1$) to their unique beneficiary center $c_n$, achieving $|B(\mu_1)| = K+1$ beneficiary matches, with $p_{K+2}$ remaining unmatched.
    
    \item A \textbf{max-elig then max-bene matching} $\mu_2$ must first maximize total matches. To match all $K+2$ parents, it must assign $p_{K+2}$ to $c_1$ (its only eligible center), which displaces $p_1$ to $c_2$, which displaces $p_2$ to $c_3$, and so on, creating a chain reaction. The resulting matching assigns $p_{K+1}$ to $c_{K+2}$ and achieves $|E(\mu_2)| = K+2$ total matches but $|B(\mu_2)| = 0$ beneficiary matches.
\end{itemize}

The difference in beneficiary matches is $|B(\mu_1)| - |B(\mu_2)| = K+1$. Since $K$ can be chosen arbitrarily large, this demonstrates that the lexicographic approach of first maximizing eligible matches can sacrifice an unbounded number of beneficiary matches.
\end{example}

\subsection{\autoref{common bene set}}\label{appendix A.2}
\begin{proposition}\label{common bene set}
Given any instance, suppose there exists a common beneficiary parent set \(\mathcal{P}_B\subseteq\mathcal{P}\) such that, for every center \(c\), \(B_c=E_c\cap\mathcal P_B\). 
Then the frontier consists of a single point. That is, there exists a matching that simultaneously maximizes total matches and beneficiary matches.
\end{proposition}

\begin{proof}
Let $\mu^\star$ be a matching that maximizes the number of beneficiary matches. A classical result on transversal matroids states that the family of subsets of parents that can be matched forms a matroid (see, e.g., \citet[Chapter~39]{schrijver2003combinatorial} or \citet[Chapter~1]{oxley2011matroid}). Hence every independent set can be extended to a basis. Therefore, the set of parents matched by $\mu^\star$ can be extended to a maximum-cardinality matchable set. Equivalently, there exists a maximum-cardinality matching $\widehat\mu$ that matches every parent already matched under $\mu^\star$.

Since beneficiary status depends only on the parent, every beneficiary matched by $\mu^\star$ remains matched under $\widehat\mu$. Thus
\[
|B(\widehat\mu)|\ge |B(\mu^\star)|.
\]
Since $\mu^\star$ maximizes the number of beneficiary matches, equality must hold. Hence $\widehat\mu$ simultaneously maximizes total matches and beneficiary matches, which completes the proof.
\end{proof}

\subsection{Proof of \autoref{disjoint cycle}}\label{proof of  disjoint cycle}
To prove \autoref{disjoint cycle}, we need the following lemma:

\begin{lemma}\label{Matroid on frontier}
    Given a matching \(\hat{\mu}\) corresponding to \(f_2=(e_2,b_2)\) on the frontier, for any $f_1=(e_1,b_1)$ with $e_1>e_2$ on the frontier, there exists a matching \(\mu^*\) corresponding to \(f_1\)  such that any parent matched in \(\hat{\mu}\) is still matched in \(\mu^*\).
\end{lemma}

\begin{proof}
    Let $\mu$ be any undominated matching corresponding to frontier point $f_1 = (e_1, b_1)$ , with $e_1-e_2=k \in \mathbb{N}$. We will construct a matching \(\mu^*\) from \(\mu\) such that any parent matched in \(\hat{\mu}\) is still matched in \(\mu^*\) and \(\mu^*\) corresponds to \(f_1\). 

Let $A$ and $D$ be the sets of matched parents in $\hat{\mu}$ and $\mu$, respectively. We partition $A$ into three disjoint sets $A_1, A_2, A_3$ such that:
\[
A_1 = \{ p \in A : \mu(p) = \hat{\mu}(p) \}, \quad 
A_2 = \{ p \in A : \mu(p) \neq \emptyset, \mu(p) \neq \hat{\mu}(p) \},\;\text{and} \]
\[A_3 = \{ p \in A : \mu(p) = \emptyset \}
.\]
Intuitively, we want to replace the parents in $\mu$ that are only matched in $\mu$ ($p\in D\setminus A$) with parents that are only matched in $\hat{\mu}$ ($p\in A_3$). With eligibility constraints, we cannot do this one-to-one exchange directly. So instead, we first match all the parents in $A_3$, then unmatch parents in $D\setminus A$ to restore the $(e_1,b_1)$ requirement. See the following diagram.
\[\mu \xrightarrow[\text{step 1}]{\text{Match all parents in }A_3}\mu'\xrightarrow[\text{step 2}]{\text{Unmatch parents in $D\setminus A$}}\mu^*.\]

\textbf{Step 1.} 
For each parent $p_0 \in A_3$, we construct a sequence starting with $p_0$, where each subsequent parent is matched in $\mu$ to the previous parent's seat in $\hat{\mu}$, until the sequence ends with an unmatched seat in $\mu$ or an unmatched parent in $\hat{\mu}$. Formally, the process is as follows.

 As $\mu$ is undominated, seat $\hat{\mu}(p_0)$ must be occupied in $\mu$ by some parent in $D$, let $p_1=\mu(\hat{\mu}(p_0))$ and add $p_1$ to the sequence. If $\hat{\mu}(p_1)=\emptyset$, then terminate the sequence with $p_1\in D\setminus A$. Else, if $\mu(\hat{\mu}(p_1))=\emptyset$, then terminate the sequence with $p_1\in A_2$; if $\mu(\hat{\mu}(p_1))\neq \emptyset$, add $p_2=\mu(\hat{\mu}(p_1))$ to the sequence and continue this process.
 
Since the set of seats is finite and all seats are visited at most once, the process ends with a finite sequence $(p_0, p_1, \ldots, p_N)$ satisfying: for every $1 \le m \le N$, $\mu(p_m) = \hat{\mu}(p_{m-1})$; all intermediate parents $p_n$ for $1 \le n < N$ belong to $A_2$; and the terminal parent $p_N$ is either in $D\setminus A$ with $\hat{\mu}(p_N)=\emptyset$, or in $A_2$ with $\mu(\hat{\mu}(p_N))=\emptyset$. We call the latter case a \textbf{type 1} sequence, and the former a \textbf{type 2} sequence.

Given any sequence $\mathbf{s} = (p_0, p_1, \ldots, p_N)$,
define the process of rematching the parents in $\mathbf{s}$ to their seats in $\hat{\mu}$ and keeping other parents' seats in $\mu$ unchanged as \(R_s(\mu)\). Concretely, for any given sequence $\mathbf{s} = (p_0, p_1, \ldots, p_N)$,

\[
R_s(\mu)(p) =
\begin{cases}
\hat{\mu}(p), & \text{if } p \in \mathbf{s}, \\
\mu(p), & \text{if } p \notin \mathbf{s}.
\end{cases}
\]
The new matching $R_s(\mu)$ is eligible and satisfies the quota constraint. Note that for any sequence $\textbf{s}$ of type 1, we have $
\{p\in \mathcal{P}: p \text{ is matched in }R_s(\mu)\}=\{p\in \mathcal{P}: p \text{ is matched in }\mu\}\cup\{p_0\}$ and thus $|E(R_s(\mu))| = |E(\mu)| + 1$, which combined with the non-domination of $\mu$ means $|B(R_s(\mu))| < |B(\mu)|$. 

For any $\textbf{s}'$ of type 2, $\{p\in \mathcal{P}: p \text{ is matched in }R_{s'}(\mu)\}=\{p\in \mathcal{P}: p \text{ is matched in }\mu\}\cup\{p_0\}\setminus\{p_N\}$ and thus $|E(R_{s'}(\mu))| = |E(\mu)|$. We claim that $|B(R_{s'}(\mu))| = |B(\mu)|$. Suppose not. If $|B(R_{s'}(\mu))| > |B(\mu)|$, then $\mu$ is dominated by $R_{s'}(\mu)$. If instead $|B(\mu)| > |B(R_{s'}(\mu))|$, rematch the parents in $\textbf{s}'$ to their seats in $\mu$ and keep other parents' seats in $\hat{\mu}$ unchanged---the new matching will dominate $\hat{\mu}$.

By construction, the sequences initialized by distinct parents in $A_3$ are disjoint. Suppose there are $k_1$ type 1 sequences and $k_2$ type 2 sequences, then $k_1 + k_2 = |A_3|$. Denote the sequences as $\mathbf{s}_1,\mathbf{s}_2,\cdots,\mathbf{s}_{|A_3|}$.
By rematching all the parents in all the sequences to their seats in $\hat{\mu}$ and keeping other parents' seats in $\mu$ unchanged, we obtain a new matching \(\mu'=R_{s_1}(R_{s_2}(\cdots (R_{s_{|A_3|}}(\mu))))\).

Let $D'$ be the set of matched parents in $\mu'$ and $A'_2 = \{ p \in A : p \text{ belongs to some sequence} \}$, then
\begin{enumerate}
    \item All parents in $A$ are matched in $\mu'$: $A \subset D' \subseteq D\cup A_3$, and \(| D'\setminus A |=|D'|-|A| =k_1+|D|-|A|=k_1+k\).
    \item $D'\setminus A\subseteq D\setminus A$.
    \item Only parents in $A_2 \setminus A'_2$ and $D' \setminus A$ get different seats (including $\emptyset$) in $\mu'$ than in $\hat{\mu}$.
    \item $|E(\mu')|-|E(\mu)|=k_1$ and $|B(\mu')|<|B(\mu)|$. Let $k_3=|B(\mu)|-|B(\mu')|$, then $k_3\ge k_1$.
\end{enumerate}

\textbf{Step 2: }
With $\mu'$ and \(\hat{\mu}\), for each parent \(p'_0\in D'\setminus A\), we can find a disjoint cycle that is also a unit eligibility improving cycle in $G_{\hat{\mu}}$. We construct the cycle by constructing a sequence like in Step 1.

Start from any parent $p'_0$ in $D' \setminus A$. As $\hat{\mu}$ is undominated, seat $\mu'(p'_0)$ must be taken by some parent under $\hat{\mu}$. Let $p'_1=\hat{\mu}(\mu'(p'_0))$. As all the parents in $A$ are matched in $\mu'$, $\mu'(p'_1)\neq\emptyset$. If $\hat{\mu}(\mu'(p'_1))=\emptyset$, then terminate the sequence; else, let $p'_2=\hat{\mu}(\mu'(p'_1))$ and continue this process.

After the above process, we have a sequence $(p'_0,p'_1,...,p'_N)$ where $\mu'(p'_m)=\hat{\mu}(p'_{m+1})$ for all $0\leq m<N$ and $\hat{\mu}(\mu'(p'_N)) =\emptyset$. Each parent $p'_m$ with $1\leq m\leq N$ belongs to $A_2 \setminus A'_2$, since $\mu'(p'_m)\neq\hat{\mu}(p'_m)$ and $\hat{\mu}(p'_m)\neq\emptyset $. 

Given any sequence generated above $(p'_0,p'_1,...,p'_N)$, we construct a unit eligibility improving cycle in $G_{\hat{\mu}}$ by linking parent $p'_{m}$ to seat $\mu'(p'_{m})$ and seat $\mu'(p'_m)$ to parent $p'_{m+1}$ for $0\leq m\leq N-1$, then linking parent $p'_N$ to seat $\mu'(p'_N)$ and seat $\mu'(p'_N)$ to parent $p'_0$.  
The unit eligibility improving cycles constructed from different sequences are disjoint. Denote these cycles as \(\textbf{c}_1,\textbf{c}_2,\dots \textbf{c}_{k+k_1}\), ordered such that \(\Delta_{\textbf{c}_1}\le \Delta_{\textbf{c}_2}\le \dots\leq \Delta_{\textbf{c}_{k+k_1}}\). Moreover, we have $\Delta_{\mathbf{c}_1}>0$; otherwise, applying $\textbf{c}_1$ to $\hat{\mu}$ would yield a matching that dominates $\hat{\mu}$. Therefore, all these disjoint unit eligibility improving cycles have positive beneficiary loss.

We claim that \(\sum_{i=1}^{k_1}\Delta_{{\textbf{c}_i}}=k_3\). We prove this by contradiction, considering two cases. 
\begin{itemize}
    \item[Case 1.] Suppose  \(\sum_{i=1}^{k_1}\Delta_{{\textbf{c}_i}}>k_3\). If we rematch the seats involved in these $k_1$ cycles to the parents they are matched to (or $\emptyset$) in $\hat{\mu}$ while keeping parents matched to other seats matched as in $\mu'$, the resulting matching would have $e_1$ eligible matches and $b_1-k_3+\sum_{i=1}^{k_1}\Delta_{{\textbf{c}_i}}>b_1$ beneficiary matches, which dominates $\mu$, a contradiction. 
    \item[Case 2.] Suppose \(\sum_{i=1}^{k_1}\Delta_{\mathbf{c}_i} < k_3\).

     First, rematch the seats involved in all type 1 sequences to the parents they are matched to (or \(\emptyset\)) in \(\mu\) and keep other seats’ parents in \(\hat{\mu}\).
    Denote the resulting matching as \(\mu_t\).

    Second, rematch the seats involved in the cycles \(\mathbf{c}_1,\mathbf{c}_2,\dots,\mathbf{c}_{k_1}\) to the parents they are matched to (or \(\emptyset\)) in \(\mu'\) and keep all other seats’ parents in \(\mu_t\). Note that this second rematch does not affect any seat or parent that was modified in the first rematch, because the sequences constructed in Step 1 and the unit eligibility improving cycles are disjoint: 
    if \( p \in A \) lies in some cycle \(\mathbf{c}\), then by the generating process of \(\mathbf{c}\), \( \mu'(p)\neq\hat{\mu}(p)\); 
    if \( p \) belongs to some sequence \(\mathbf{s}\), then by the construction of \(\mu'\), \( \mu'(p)=\hat{\mu}(p)\).

    The resulting matching has \(e_2 - k_1 + k_1 = e_2\) eligible matches and \(b_2 + k_3 - \sum_{i=1}^{k_1}\Delta_{\mathbf{c}_i} > b_2\) beneficiary matches, thus dominating \(\hat{\mu}\), a contradiction. 
\end{itemize}

Now we construct \(\mu^*\). Rematch all the seats involved in the cycles \(\textbf{c}_1,\dots,\textbf{c}_{k_1}\) to the parents they are matched to in $\hat{\mu}$ and keep parents matched to other seats matched as in $\mu'$. We call this process the reverse application of the unit eligibility improving cycles. We claim the resulting matching $\mu^*$ satisfies the requirements. Since \(\sum_{i=1}^{k_1}\Delta_{{\textbf{c}_i}}=k_3\), we have $|B(\mu^*)|=|B(\mu)|=b_1$. The fact that $|E(\mu')|-|E(\mu)|=k_1$ and that $\mu^*$ is obtained by reversely applying $k_1$ unit eligibility improving cycles implies $|E(\mu^*)|=|E(\mu)|=e_1$. To show that $\mu^*$ matches all the matched parents in $\hat{\mu}$, note that $\mu'$ has all the parents in $A$ matched, and the reverse application of unit eligibility improving cycles in Step 2 only unmatches parents in $D'\setminus A$.
\end{proof}

\begin{proof}[Proof of \autoref{disjoint cycle}]
In this proof, we use some notation from the proof of \autoref{Matroid on frontier}. 

By \autoref{Matroid on frontier}, there exists a matching \(\mu^*\) corresponding to \(f_1\) such that any matched parent in \(\hat{\mu}\) is still matched in \(\mu^*\). Partition $\mathcal{C}$ into five subsets $C_1,\cdots, C_5$, separately containing: seats of parents in the $k_1$ type-1 sequences; seats in the $k$ disjoint unit eligibility improving cycles that have not been used in constructing $\mu^*$; seats in the $k_1$ disjoint unit eligibility improving cycles that have been used in constructing $\mu^*$; seats of parents in the $k_2$ type-2 sequences; and all other seats. The partition is shown below:

\[
\text{Partition } \mathcal{C} \text{ into }
\left\{
\begin{aligned}
C_1 &:~ \text{seats of parents in the } k_1 \text{ type-1 sequences},\\[4pt]
C_2 &:~
\begin{aligned}[t]
&\text{seats in the } k \text{ disjoint unit eligibility improving cycles}\\[-1pt]
&\text{not used in constructing } \mu^*,
\end{aligned}\\[4pt]
C_3 &:~
\begin{aligned}[t]
&\text{seats in the } k_1 \text{ disjoint unit eligibility improving cycles}\\[-1pt]
&\text{used in constructing } \mu^*,
\end{aligned}\\[4pt]
C_4 &:~ \text{seats of parents in the } k_2 \text{ type-2 sequences},\\[4pt]
C_5 &:~ \text{other seats.}
\end{aligned}
\right.
\]

By construction, the seats in $C_1,C_3,C_4$ are all matched to the same parents in $\mu^*$ and $\hat{\mu}$, while the seats in $C_2$ and $C_5$ are all matched to the same parents (or are both unmatched) in $\mu^*$ and $\mu$. 

Apply the $k$ disjoint unit eligibility improving cycles that have not been used in constructing $\mu^*$, proved in \autoref{Matroid on frontier} to have positive beneficiary loss, to $\hat{\mu}$, and denote the resulting matching as $\mu_3$. Then $\mu_3$ matches seats in $C_1,C_3,C_4,C_5$ to the same parents as in $\hat{\mu}$, and seats in $C_2$ to the same parents as in $\mu$. We will show that $\mu_3$ corresponds to $(e_1,b_1)$.

As $\mu^*$ corresponds to $(e_1,b_1)$, and $\mu^*$ and $\mu_3$ differ only in the matches of seats in $C_5$, it suffices to show that  $\mu$ and $\hat{\mu}$ have the same numbers of eligible and beneficiary matches when restricted to seats in $C_5$. 

For simplicity, we will refer to seats of parents in sequences as seats belonging to those sequences. 

Consider any parent involved in a cycle or sequence.  If this parent is matched in both $\mu$ and $\hat{\mu}$, then in each matching it must be matched to a seat belonging to the same cycle or sequence. Additionally, parents matched only in $\hat{\mu}$ belong to sequences, while parents matched only in $\mu$ belong to cycles. As a consequence, any parent matched to a seat in $C_5$ under $\mu$ must also be matched to a seat in $C_5$ under $\hat{\mu}$. Therefore, when restricted to $C_5$, both matchings have the same number of eligible matches and match the same set of parents. 

We claim they also have the same number of beneficiary matches. Suppose for contradiction, without loss of generality, $\mu$ has more beneficiary matches than $\hat{\mu}$ when restricted to $C_5$. Consider the matching obtained by rematching the seats in $C_5$ according to $\mu$ while keeping all other seats' matches as in $\hat{\mu}$. This new matching would dominate $\hat{\mu}$, contradicting its non-domination.  Therefore, $\mu$ and $\hat{\mu}$ have the same number of beneficiary matches when restricted to $C_5$, which implies that $\mu_3$ corresponds to $(e_1,b_1)$, completing the proof.
\end{proof}

\subsection{Proof of \autoref{slope change points}}\label{proof of  slope change points}
\begin{proof}
Note that given a weight,  a dominated matching will have a smaller weight sum than the matching that dominates it. Thus only undominated matchings can be maximum-weight matchings in any iteration of the first loop of \autoref{alg:RHA}. It suffices to show that in each iteration, the stated undominated matching has a higher weight sum than all other undominated matchings.

\medskip
\noindent\textbf{Proof of Claim 1.}

Consider the first and the last iterations of the first loop of \autoref{alg:RHA}. In the first iteration ($k=1$), the weights are
\[
w_{\mathrm{elig}} = 1,\qquad w_{\mathrm{bene}} = 1 + \frac{1}{1} + \frac{1}{n^2} = 2 + \frac{1}{n^2}.
\]

The weight sum of any max-bene then max-elig matching $\mu$ is
\[
\left(e-K\right)+\left(b+\sum^K_{i=1}\Delta b_i\right)\left(1 + \frac{1}{n^2}\right)
.\]

Meanwhile, the weight sum of any other undominated matching corresponding to the point \((e-T,b+\sum^T_{i=1}\Delta b_i)\) for \(0\le T<K\) is 
\[(e-T)+(b+\sum^T_{i=1}\Delta b_i)(1 + \frac{1}{n^2})
.\]

Since \(\Delta b_i\ge1\) for all \(i\in \{1,\dots K\}\) by \autoref{slope of the frontier}, the difference is

\[(\sum^K_{i=T+1}\Delta b_i)(1+\frac{1}{n^2})-(K-T)>0.\]

 Therefore, any max-bene then max-elig matching has a higher weight sum than other undominated matchings (under this weight), so one of the max-bene then max-elig matchings will be chosen in the first iteration.

In the last iteration ($k=n$) the weights are
\[
w_{\mathrm{elig}} = 1,\qquad w_{\mathrm{bene}} = 1 + \frac{1}{n} + \frac{1}{n^2}.
\]

The weight sum of any max-elig then max-bene matching is 

\[e+\left(\frac{1}{n} + \frac{1}{n^2}\right)b.\]

The weight sum of any other undominated matching corresponding to  \((e-T,b+\sum^T_{i=1}\Delta b_i)\) for \(1\le T\le K\) is 

\[\left(e-T\right)+\left(b+\sum^T_{i=1}\Delta b_i\right)\left(\frac{1}{n} + \frac{1}{n^2}\right)
.\]

Since \(\sum^T_{i=1}\Delta b_i\le e-T\le n-T\) and \(T\ge1\), the difference is
\[T-\left(\sum^T_{i=1}\Delta b_i\right)\left(\frac{1}{n} + \frac{1}{n^2}\right) \ge T-(n-T)\left(\frac{1}{n} + \frac{1}{n^2}\right)=T\left(1 + \frac{1}{n} + \frac{1}{n^2}\right) - \left(1 + \frac{1}{n}\right)>0.\]

Therefore, any max-elig then max-bene matching achieves a higher weight sum than other undominated matchings (under this weight), so one of the max-elig then max-bene matchings will be chosen in the last iteration.

\medskip
\noindent\textbf{Proof of Claim 2.}

Fix a kink $f_j=(e-j,b+\sum_{i=1}^j\Delta b_i)$ on the frontier such that $\Delta b_{j}>\Delta b_{j+1}$. We show that RHA computes its $(e,b)$ values in the $\Delta b_j$-th iteration.

In the $\Delta b_j$-th iteration, the weights are
\[
w_{\mathrm{elig}} = 1, \qquad w_{\mathrm{bene}} = 1 + \frac{1}{\Delta b_j} + \frac{1}{n^2}.
\]

We show that any matching corresponding to $f_j$ has a higher weight sum than any matching corresponding to another frontier point $f_t$. We consider two cases.

\textbf{Case 1. $t<j$.} The difference in weight sums between $f_j$ and $f_t$ is 
\[
\left( \frac{1}{\Delta b_j} + \frac{1}{n^2} \right)\sum_{i=t+1}^{j} \Delta b_i - (j-t)\geq \left( \frac{1}{\Delta b_j} + \frac{1}{n^2} \right)(j-t) \Delta b_j - (j-t)=(j-t) \frac{\Delta b_j}{n^2}>0.
\]

Thus, any frontier point $f_t$ with $t<j$ will not be chosen in the $\Delta b_j$-th iteration.

\textbf{Case 2. $t>j$.} First consider $t=j+1$. The difference in weight sum between $f_j$ and $f_{j+1}$ is
\[
1 - \left( \frac{1}{\Delta b_j} + \frac{1}{n^2} \right)\Delta b_{j+1}.
\]

Since $\frac{1}{n^2} < \frac{1}{\Delta b_j^2} < \frac{1}{\Delta b_j \Delta b_{j+1}}$ and $\Delta b_j > \Delta b_{j+1}$, we have
\begin{equation}\label{appendix j+1}
    1 - \left( \frac{1}{\Delta b_j} + \frac{1}{n^2} \right)\Delta b_{j+1} 
> 1 - \left( \frac{1}{\Delta b_j} + \frac{1}{\Delta b_j \Delta b_{j+1}} \right)\Delta b_{j+1}
= \frac{\Delta b_j - \Delta b_{j+1} - 1}{\Delta b_j} \ge 0.
\end{equation}

Thus, the frontier point $f_{j+1}$ will not be chosen in the $\Delta b_j$-th iteration.

For any $t > j+1$, the difference in weight sums between $f_j$ and $f_t$ is
\[
(t-j) - \left( \frac{1}{\Delta b_j} + \frac{1}{n^2} \right) \sum_{i=j+1}^t \Delta b_i.
\]

Since $\Delta b_{j+1} \ge \Delta b_{j+2} \ge \cdots \ge\Delta b_{t}$, we have
$\sum_{i=j+1}^t \Delta b_i \le (t-j)\Delta b_{j+1}$.  Combined with \autoref{appendix j+1}, this gives
\[
(t-j) - \left( \frac{1}{\Delta b_j} + \frac{1}{n^2} \right) \sum_{i=j+1}^t \Delta b_i\geq(t-j) - \left( \frac{1}{\Delta b_j} + \frac{1}{n^2} \right)(t-j)\Delta b_{j+1} > 0.
\]

Thus, any frontier point $f_{t}$ with $t>j+1$ will not be chosen in the $\Delta b_j$-th iteration. 

Therefore, the matching corresponding to $f_j$ has the highest weight sum in the $\Delta b_j$-th iteration.

\end{proof}

\subsection{Proof of \autoref{justification of approximation}}\label{proof of justification of approximation}
\begin{proof}
    Consider the non-trivial case where a matching with beneficiary-share $\beta^*$ exists but no frontier point achieves exactly $\beta^*$. Among all matchings $\nu$ satisfying $\beta(\nu)=\beta^*$, let
    $\mu'$ be one that maximizes $|E(\nu)|$.
    Then \(\mu'\) dominates all other matchings with the same beneficiary-share but fewer total matches. To prove the proposition, it suffices to show that $\mu^*$ dominates $\mu'$. We consider two cases.

    \textbf{Case 1:} \(|E(\mu')| < |E(\mu_{BE})|\). Since \(\mu^*\) is undominated and $\mu_{BE}$ is the undominated matching with the least eligible matches, we have \(|E(\mu^*)| \ge |E(\mu_{BE})| > |E(\mu')|\). If \(\beta(\mu^*) > \beta(\mu')\), then \(\left |  B(\mu^*)\right | >\left |  B(\mu')\right | \), so \(\mu^*\) clearly dominates \(\mu'\). If instead \(\beta(\mu^*) < \beta(\mu')\), then by definition \(\mu^*\) is a max-bene then max-elig matching, implying \(|B(\mu^*)| \ge |B(\mu')|\). Combined with \(|E(\mu^*)| > |E(\mu')|\), we have that \(\mu^*\) dominates \(\mu'\).

    \textbf{Case 2:} \(|E(\mu')| \ge |E(\mu_{BE})|\). By \autoref{slope of the frontier}, there exists an undominated matching \(\mu''\) with \(|E(\mu'')| = |E(\mu')|\) and \(|B(\mu'')| > |B(\mu')|\), implying \(\beta(\mu'') > \beta(\mu') = \beta^*\). Since \(\mu^*\) respects the beneficiary-share guarantee approximately on the frontier and \(\beta(\mu^*) \ne \beta^*\), we must have \(\beta(\mu^*) > \beta^*=\beta(\mu')\) and \(|E(\mu^*)| \ge |E(\mu'')| = |E(\mu')|\). This implies $|B(\mu^*)|=\beta(\mu^*)|E(\mu^*)|>|B(\mu')|$. Therefore, \(\mu^*\) dominates \(\mu'\).

    In both cases, \(\mu^*\) dominates \(\mu'\), completing the proof.
\end{proof}

\subsection{Proof of \autoref{Respecting priority}}\label{proof of Respecting priority}
\begin{proof}
    For any eligible matching \(\mu\), define its \textbf{rank sum} as \(R(\mu)=\sum_{p:\,\mu(p)\neq\emptyset}R_{\mu(p)}(p)\), where $R_c(p)$ denotes the position of parent $p$ in center $c$'s priority order $\pi_c$ (with $R_c(p) = 1$ for the highest-priority parent).
    
    Consider any point $(e,b)$ on the frontier. Among all matchings achieving this point, the matching with minimum rank sum must respect priority. To see why, suppose matching $\mu$ achieves $(e,b)$ with minimum rank sum but violates priority: for some center $c$, we have $\mu(p) = c$ and $\mu(p') = \emptyset$ with $p' \pi_c p$. Then swapping assignments to set $\mu'(p') = c$ and $\mu'(p) = \emptyset$ would yield another matching achieving $(e,b)$ with strictly lower rank sum, contradicting minimality.

    Therefore, our mechanism operates as follows: First, identify the frontier point $(e^*, b^*)$ corresponding to a matching that respects the beneficiary-share guarantee approximately on the frontier. Then, among all matchings achieving $(e^*, b^*)$, select one minimizing the rank sum. This matching respects both priority and the beneficiary-share guarantee approximately on the frontier.
\end{proof}

\subsection{Proof of \autoref{Strategy-proof}}\label{proof of  Strategy-proof}
\begin{proof}
For each report profile $A$, the mechanism first chooses the frontier point prescribed by approximate beneficiary-share compliance: if some frontier point has beneficiary share at least $\beta^*$, it chooses the one whose beneficiary share is closest to $\beta^*$ from above; otherwise, it chooses the max-bene-then-max-elig endpoint. Among all matchings corresponding to the chosen frontier point, a set of matchings minimize the priority-rank sum. By \autoref{Respecting priority}, a mechanism that always chooses a matching from that set respects priority. Let $\varphi$ be a mechanism that always selects a matching from that set according to a strict ordering of matchings that is fixed independently of the reports. We will prove that $\varphi$ is strategy-proof.

Suppose, for contradiction, that $\varphi$ is not strategy-proof. Then there exists a parent $p_i$ with true acceptable set $T_i$ who gains from reporting $A_i'\neq T_i$. Let
$A=(T_i,A_{-i})$ be the truthful profile and
$A'=(A_i',A_{-i})$ the deviation. 
Then $\varphi(A)(p_i)=\emptyset$ and $\varphi(A')(p_i)\in T_i$.
Write $\mu=\varphi(A)$ and $\mu'=\varphi(A')$, and let $(e,b)$ and
$(e',b')$ be their respective numbers of eligible and beneficiary
matches.

We first establish eligibility. Say a matching is eligible under a report if all of its matches are eligible. Since $p_i$ is unmatched in $\mu$, changing only her report does not affect the eligibility of matches in $\mu$, so $\mu$ remains eligible under $A'$. Conversely, since $\mu'(p_i)\in T_i$ and all other reports are unchanged, every match in $\mu'$ is eligible under $A$, so $\mu'$ is eligible under $A$. Since the beneficiary relation is fixed independently of parents' reports, the numbers of eligible and beneficiary matches of each matching are the same under the two profiles.

It follows that either $(e,b)=(e',b')$, or one matching has strictly more eligible matches and strictly fewer beneficiary matches than the other. To see this, suppose instead that $e'\ge e$ and $b'\ge b$, with at least one inequality strict. Since $\mu'$ is eligible under $A$, it would dominate $\mu$, contradicting that $(e,b)$ lies on the frontier under $A$. Symmetrically, $e\ge e'$ and $b\ge b'$, with at least one inequality strict, is impossible because $\mu$ is eligible under $A'$.

\medskip
\noindent\textbf{Case 1: $(e,b)\neq(e',b')$.}

We will consider the case where $e'<e$ and $b'>b$. The opposite case holds by a symmetric version of the following argument which only uses the fact that $\mu$ and $\mu'$ are eligible under both profiles and each satisfies the beneficiary share guarantee approximately on the frontier under its corresponding profile.

Suppose first that $(e,b)$ satisfies the beneficiary-share target under $A$, so $b/e\ge\beta^*$. Since $\mu$ remains eligible under $A'$, the maximum number of eligible matches attainable under $A'$ is at least $e$. Moreover, $(e',b')$ is a frontier point under $A'$ and $e'<e$. Hence $e$ lies between the eligible-match counts of the two frontier endpoints under $A'$. Since the frontier is exhaustive in eligible matches by \autoref{slope of the frontier}, the frontier under $A'$ contains a point $(e,\widetilde b)$ with $\widetilde b\ge b$. Hence
\[
\frac{\widetilde b}{e}\ge \frac{b}{e}\ge\beta^*.
\]
Moreover, by non-domination (\autoref{def of frontier}) and the frontier structure in \autoref{slope of the frontier}, beneficiary matches strictly decrease as eligible matches increase along distinct frontier points. Therefore beneficiary share also strictly decreases with the number of eligible matches whenever the beneficiary count is positive. Since $(e',b')$ and $(e,\widetilde b)$ are frontier points under $A'$ and $e'<e$,
\[
\frac{b'}{e'}>\frac{\widetilde b}{e}\ge\beta^*.
\]
Thus $(e,\widetilde b)$ also satisfies the target but has a beneficiary share closer to $\beta^*$ than $(e',b')$, contradicting the choice of $(e',b')$ under $A'$.

Suppose instead that $(e,b)$ is chosen because the target is unattainable on the frontier under $A$. Then $(e,b)$ is the max-bene-then-max-elig endpoint, so $b$ is the largest attainable number of beneficiary matches under $A$. But $\mu'$ is eligible under $A$ and $b'>b$, a contradiction. Hence $(e,b)$ and $(e',b')$ cannot be distinct.

\medskip
\noindent\textbf{Case 2: $(e,b)=(e',b')$.}

In this case, $\mu$ and $\mu'$ are eligible under both report profiles and correspond to the same frontier point. Let $R(\nu)$ denote the priority-rank sum of a matching $\nu$. Since $\mu$ is chosen under $A$ while $\mu'$ is also eligible and corresponds to the same frontier point, $R(\mu)\le R(\mu')$. Likewise, $R(\mu')\le R(\mu)$ under $A'$. Hence
\[
R(\mu)=R(\mu').
\]

The choice between $\mu$ and $\mu'$ must therefore be determined by the same report-independent strict tie-breaking order under both profiles. That is, either $\mu$ ranks higher than $\mu'$ in the tie-breaking order or the opposite. It cannot be the case that one is chosen under $A$ and another is chosen under $A'$, a contradiction.

Both cases yield a contradiction. Therefore no parent who is unmatched under truthful reporting can obtain an assignment in her true acceptable set by misreporting her family-side eligibility information. Hence $\varphi$ is strategy-proof with respect to these reports.
\end{proof}

\subsection{Proof of \autoref{thrm PI imposible}}\label{proof of thrm PI imposible}
\begin{proof}
    Consider a problem where $\mathcal{P} = \{p_1,p_2,p_3,p_4,p_5,p_6\}$, \(\mathcal{C} = \{c_1,c_2,c_3,c_4,c_5\}\), each with capacity \(1\), and \(\beta^* = 0.2\). The eligibility and beneficiary relations are given in the following table, where boxed entries indicate beneficiaries of the corresponding centers:

    \[
    \begin{array}{c|ccccc}
        & c_1 & c_2 & c_3 & c_4 & c_5 \\
        \hline
         & \boxed{p_1} & p_2 & p_3 & \boxed{p_6} & p_1 \\
         & p_2 & p_3 & p_5 & p_4 & \\
    \end{array}
    \]
    
    Consider \(X = \{p_1,p_2,p_3,p_4,p_5\}\). The undominated matchings for the subproblem $(\mathcal{C},X,\textbf{q},\mathcal{E}, \mathcal{B},\beta^*)$ with parent set $X$ are: 
  \begin{align*}
    b = 1,\ e = 4,\ \beta = 0.25 &: 
    \{(c_1,\boxed{p_1}), (c_2,p_2), (c_3,p_3), (c_4,p_4), (c_5,\emptyset)\}, \\
    &\quad \text{} \{(c_1,\boxed{p_1}), (c_2,p_2), (c_3,p_5), (c_4,p_4), (c_5,\emptyset)\}, \\
    &\quad \text{and } \{(c_1,\boxed{p_1}), (c_2,p_3), (c_3,p_5), (c_4,p_4), (c_5,\emptyset)\}. \\
    b = 0,\ e = 5,\ \beta = 0 &: 
    \{(c_1,p_2), (c_2,p_3), (c_3,p_5), (c_4,p_4), (c_5,p_1)\}.
\end{align*}

    Since $\beta^* = 0.2$ and the first three matchings achieve $\beta = 0.25 > \beta^*$, any mechanism respecting the beneficiary-share guarantee approximately on the frontier must select one of these three matchings. Therefore, $C(X) \in \{\{p_1,p_2,p_3,p_4\}, \{p_1,p_2,p_4,p_5\}, \{p_1,p_3,p_4,p_5\}\}$.

    Now consider the full parent set $\mathcal{P} = X \cup \{p_6\}$. The frontier becomes:
    \begin{align*}
    b = 2,\ e = 4,\ \beta = 0.5 &: 
    \{(c_1,\boxed{p_1}), (c_2,p_2), (c_3,p_3), (c_4,\boxed{p_6}), (c_5,\emptyset)\}, \\
    &\quad  \{(c_1,\boxed{p_1}), (c_2,p_2), (c_3,p_5), (c_4,\boxed{p_6}), (c_5,\emptyset)\}, \\
    &\quad \text{and } \{(c_1,\boxed{p_1}), (c_2,p_3), (c_3,p_5), (c_4,\boxed{p_6}), (c_5,\emptyset)\}. \\
    b = 1,\ e = 5,\ \beta = 0.2 &: 
    \{(c_1,p_2), (c_2,p_3), (c_3,p_5), (c_4,\boxed{p_6}), (c_5,p_1)\}.
\end{align*}

    Since the matching with $\beta = 0.2$ exactly achieves $\beta^*$, the mechanism must select it. Thus $C(\mathcal{P}) = \{p_1,p_2,p_3,p_5,p_6\}$.

    However, for any possible \(C(\cdot)\), we have:
    \[
    C(\{p_1,p_2,p_3,p_4,p_5,p_6\}) \cap \{p_1,p_2,p_3,p_4,p_5\} = \{p_1,p_2,p_3,p_5\} \not\subseteq C(\{p_1,p_2,p_3,p_4,p_5\}),
    \]
    which violates substitutability, and hence path-independence.\footnote{The example in the proof actually demonstrates a violation of the \textit{substitutability } \citep[see recent discussion in][]{bando2025properties}, which states that if $X' \subseteq X$, then $C(X) \cap X' \subseteq C(X')$. In our example, we have $X \subseteq \mathcal{P}$ with $C(\mathcal{P}) \cap X = \{p_1,p_2,p_3,p_5\} \nsubseteq C(X)$ for any valid $C(X)$. Since our mechanism violates substitutability, it necessarily violates path-independence \citep{aizerman1981}.}
\end{proof}

\subsection{Proof of \autoref{prop:ce lp}}\label{proof of  prop:ce lp}
\begin{proof}
For any eligible matching $\mu$, let $x^\mu$ denote its incidence
vector, where $x^\mu_{pc}=1$ if $\mu(p)=c$ and $x^\mu_{pc}=0$
otherwise. Every integral feasible solution of \eqref{lp:ce}
corresponds to an eligible matching $\mu$ with $|E(\mu)|=e$, and
conversely.

Since $M>\sum_{p,c}R_p(c)$, the objective first maximizes the number of
beneficiary matches among matchings with exactly $e$ eligible matches
and then minimizes their aggregate preference-rank sum. Because
$(e,b)$ lies on the non-domination frontier, the maximum beneficiary
count attainable with $e$ eligible matches is $b$. Hence every integral
optimal solution $x^{\mu^\star}$ induces a matching $\mu^\star$
corresponding to $(e,b)$.

Suppose that a matching $\mu'$ corresponding to $(e,b)$ Pareto
dominates $\mu^\star$. Since both matchings have exactly $e$ matches
and every parent prefers any eligible center to remaining unmatched,
they must match the same set of parents. Every matched parent then
weakly prefers $\mu'(p)$ to $\mu^\star(p)$, with at least one strict
preference. Therefore,
$\sum_{p,c}R_p(c)x^{\mu'}_{pc}
<
\sum_{p,c}R_p(c)x^{\mu^\star}_{pc}$.
Because $|B(\mu')|=|B(\mu^\star)|=b$, the two solutions have the same
$M$-term, so $x^{\mu'}$ attains a strictly larger objective value than
$x^{\mu^\star}$, a contradiction. Thus, $\mu^\star$ is constrained
efficient at $(e,b)$.

Finally, \eqref{lp:ce} is equivalent to a minimum-cost flow problem.
Connect the source to each parent with unit capacity and zero cost;
connect parent $p$ to center $c$ whenever $p\in E_c$, with unit
capacity and cost $R_p(c)-M\mathbf{1}\{p\in B_c\}$; and connect each
center $c$ to the sink with capacity $q_c$ and zero cost. Requiring
exactly $e$ units of flow yields precisely the feasible region of
\eqref{lp:ce}. Since all capacities and the required flow are integral,
the problem admits an integral optimum and can be solved in polynomial
time \citep{ahuja1993network}.
\end{proof}

\subsection{Proof of \autoref{aligned efficiency}}\label{proof of aligned efficiency}
\begin{proof}
    Fix any $(e,b)\in F$. Let
\[
\mu^*
\in
\underset{\nu\text{ corresponding to }(e,b)}{\arg\min}
\sum_{p:\,\nu(p)\neq\emptyset}R_p(\nu(p)).
\] We claim \(\mu^*\) is Pareto efficient. Suppose not, then there exists some matching \(\mu'\) that Pareto dominates \(\mu^*\). As preferences are aligned with beneficiary status, we get \(|E(\mu')|\ge |E(\mu^*)|\) and \(|B(\mu')|\ge |B(\mu^*)|.\) So it must be the case that \(\mu'\) also corresponds to \((e,b)\) otherwise \(\mu^*\) would be dominated. This means \(\mu'\) cannot have a smaller preference rank sum than \(\mu^*\) by definition, which contradicts that \(\mu^*\) is Pareto dominated by \(\mu'\).
\end{proof}

\end{document}